\documentclass[twocolumn]{aastex61}
\usepackage{amsmath,amstext}
\usepackage[T1]{fontenc}
\usepackage{apjfonts} 
\usepackage[figure,figure*]{hypcap}
\usepackage{gensymb}
\usepackage{ amssymb }
\usepackage{float}

\shorttitle{The nuclear star cluster of the Sgr dSph galaxy}
\shortauthors{M. Alfaro-Cuello et al.}

\begin{document}

\title{A deep view into the nucleus of the Sagittarius Dwarf Spheroidal Galaxy with MUSE.\\
 I. Data and stellar population characterization }

\author{M. Alfaro-Cuello}
\affiliation{Max-Planck-Institut f\"ur Astronomie, K\"onigstuhl 17, 69117 Heidelberg, Germany.}

\author{N. Kacharov}
\affiliation{Max-Planck-Institut f\"ur Astronomie, K\"onigstuhl 17, 69117 Heidelberg, Germany.}

\author{N. Neumayer}
\affiliation{Max-Planck-Institut f\"ur Astronomie, K\"onigstuhl 17, 69117 Heidelberg, Germany.}

\author{N. L\"utzgendorf}
\affiliation{European Space Agency, c/o STScI, 3700 San Martin Drive, Baltimore, MD 21218, USA.} 

\author{A.C. Seth}
\affiliation{Department of Physics and Astronomy, University of Utah, Salt Lake City, UT 84112, USA.}

\author{T. B\"oker}
\affiliation{European Space Agency, c/o STScI, 3700 San Martin Drive, Baltimore, MD 21218, USA.} 

\author{S. Kamann}
\affiliation{Astrophysics Research Institute, Liverpool John Moores University, 146 Brownlow Hill, Liverpool L3 5RF, United Kingdom.}

\author{R. Leaman}
\affiliation{Max-Planck-Institut f\"ur Astronomie, K\"onigstuhl 17, 69117 Heidelberg, Germany.}

\author{G. van de Ven}
\affiliation{Department of Astrophysics, University of Vienna, T\"urkenschanzstrasse 17, 1180 Wien, Austria.}

\author{P. Bianchini}
\affiliation{Observatoire astronomique de Strasbourg, CNRS, UMR 7550, F-67000 Strasbourg, France.}

\author{L. L. Watkins}
\affiliation{Space Telescope Science Institute, 3700 San Martin Drive, Baltimore, MD 21218, USA.}
\affiliation{European Southern Observatory, Karl-Schwarzschild-Str. 2, 85748 Garching, Germany.}

\author{M. Lyubenova}
\affiliation{European Southern Observatory, Karl-Schwarzschild-Str. 2, 85748 Garching, Germany.}

\begin{abstract}

The center of the Sagittarius dwarf spheroidal galaxy (Sgr dSph) hosts a nuclear star cluster, M54, which is the only galaxy nucleus that can be resolved into individual stars at optical wavelengths.  It is thus a key target for understanding the formation of nuclear star clusters and their relation to globular clusters.  We present a large Multi-Unit Spectroscopic Explorer (MUSE) data set that covers M54 out to $\sim$2.5 half-light radius, from which we extracted the spectra of $\sim$6\,600 cluster member stars. We use these data in combination with HST photometry to derive age and metallicity for each star. The stellar populations show a well defined age-metallicity relation, implying an extended formation history for the central region of Sgr dSph. We classify these populations into three groups, all with the same systemic velocity: young metal rich (YMR; 2.2\,Gyr, \mbox{[Fe/H]$=-0.04$}); intermediate age metal rich (IMR; 4.3\,Gyr, \mbox{[Fe/H]$=-0.29$}); and old metal poor (OMP; 12.2\,Gyr, \mbox{[Fe/H]$=-1.41$}). The YMR and OMP populations are more centrally concentrated than the IMR population, which are likely stars of the Sgr dSph. We suggest the OMP population is the result of accretion and merging of two or more old and metal poor globular clusters dragged to the center by dynamical friction. The YMR is consistent with being formed by in situ star formation in the nucleus. The ages of the YMR population suggest that it may have been triggered into forming when the Sgr dSph began losing its gas during the most recent interaction with the Milky Way, $\sim$3\,Gyr ago. 
\end{abstract}


\section{Introduction}

Nuclear star clusters (NSCs) are the densest stellar systems in the Universe \citep{Walcher2005, Misgeld_Hilker2011, Norris2014} with masses from 10${^6}$-10${^8}$\,M$_{\odot}$ and half-light radii of about 1-10\,pc \citep{Georgiev_Boker2014, Georgiev2016}.
High resolution Hubble Space Telescope (HST) imaging of nearby galaxies has revealed that NSCs are found in $\gtrsim$70\% of galaxies across the Hubble sequence \citep{Phillips1996, Carollo1998, Boker2002, Boker2004, Cote2006, Turner2012, Georgiev_Boker2014} covering a wide range of galaxy stellar masses with an occupation fraction peaking at galaxy masses in the range of 10$^{8-10}~M_\odot$ \citep{Georgiev2009, Ordenes-Briceno2018, sanchez2018}. 
In addition, NSCs are known to co-exist with supermassive black holes \citep[SMBH,][]{Seth2008, Neumayer_Walcher2012}, as observed in the Milky Way \citep{Genzel2010, Schoedel2014, schoedel2014b, feldmeier2014}.

NSCs may also be relevant for the origin of the most massive globular clusters (GCs), which is still a debated question. Many of the most massive clusters in the Milky Way and M31 show significant spreads in metallicity \citep{Meylan2001,Fuentes-Carrera2008,Willman-Strader2012,Bailin2018}, and the most massive Milky Way GC, $\omega$~Cen \citep[2.5$\times$10$^6~M_\odot$,][]{vandeVen2006}, seems to have age spreads as well \citep{Hilker2004,Villanova2014}. Terzan\,5 is another multi-component cluster that resides in the bulge of the Milky Way. It displays a high spread in iron \citep{Massari2014} and age \citep{Ferraro2016, Origlia2019}, with a chemical composition similar to the Galactic bulge.
This led to the suggestion that they are former NSCs of dwarf galaxies that have been accreted \citep{Boker2008,daCosta2016}, as nuclear star clusters are known to have extended star formation histories \citep[e.g.][]{Walcher2005,Kacharov2018}. 
Metallicity spreads alone could also be due to self-enrichment during formation \citep{Bailin2018}, however age spreads are likely a strong indicator of a stripped NSCs. \citet{Pfeffer2014} predict, via cosmological simulations of the Milky Way, that between one and three GCs with masses higher than 10$^5~$M$_\odot$ are tidally striped nuclei, with a high likelihood of remnants above 10$^6~$M$_\odot$.  

The most massive compact stellar systems are often referred to as Ultracompact Dwarfs (UCDs), which blend smoothly in mass and size with the massive GCs with no clear dividing line between them.  For massive UCDs, above 10$^7~M_\odot$, a large fraction do appear to be stripped nuclei, as they host SMBHs that make up a large fraction of their mass \citep{Seth2014,Ahn2017,Ahn2018,Afa2018} or have an extended star formation history \citep{Norris2015}.  At masses below 10$^7~M_\odot$, no equivalent evidence has been found outside of $\omega$~Cen's extended star formation history, but stripped NSCs at these masses are expected to be quite common \citep{Georgiev_Boker2014,Pfeffer2014}.  

In this context, the following questions arise: how many of the massive GCs in our Milky Way are former nuclei of low-mass galaxies?  What are the detailed properties (morphology, stellar ages, metallicity, and kinematics) we would expect these systems to have?  

To address these questions, we focus here on an ideal target in the nucleated Sagittarius dwarf spheroidal galaxy \citep{Ibata1994} - hereafter Sgr dSph. This galaxy is currently being disrupted by the tidal field of the Milky Way \citep{Ibata1997}, leaving visible streams as evidence.  The original luminosity of the galaxy has been estimated to be $\sim$10$^8$~L$_\odot$ \citep[$M_v \sim -15$;][]{Niederste-Ostholt2012}.

The NSC of this galaxy was discovered as the globular cluster Messier~54 (NGC\,6715) long before the remainder of the galaxy was discovered.  M54 has a mass of \mbox{1.41$\pm0.02\times$10${^6}$\,M$_\odot$} \citep{Baumgardt2018}, making it the second most massive GC in the galaxy after $\omega$~Cen. It is located in the densest region of the Sagittarius stream at the photometric center of the galaxy \citep{Ibata1994, Mucciarelli2017}  at a distance of 28.4\,kpc \citep{Siegel2011}.

The Sgr dSph NSC is composed of at least two distinct populations, a metal poor and metal rich, with the metal-rich population having stars as young as $\sim$2~Gyr, while the metal-poor population is consistent with an age of $\sim$13~Gyr \citep{Monaco2005b, Siegel2007, Bellazzini2008}.  Some authors have referred to these two components as separate objects, a massive metal-poor GC (M54), and a metal-rich component associated with the galaxy (Sgr, or Sgr NS).  However, in the context of extragalactic NSCs (where these two populations could not easily be separated), and given the populations' identical photocenters \citep{Monaco2005a} and radial velocities \citep{Bellazzini2008} we prefer to think of these components as two (or more) subpopulations of the Sgr dSph NSC. 
Hence, we will refer throughout the paper to the different subpopulations in the NSC of the Sgr dSph considering their stars physical parameters (e.g metallicity, age).

\citet{Monaco2005a} found that the metal-rich subpopulation presents a cusp with a peak located in an indistinguishable position from the center of the metal-poor population, thus displaying a different radial profile in the inner parts in comparison with the rest of Sgr dSph galaxy \citep{Majewski2003}. \citet{Ibata2009} reported that the metal-poor subpopulation shows a stellar density cusp, together with a peak in the velocity dispersion. The density profiles from both subpopulations are found to be different \citep{Monaco2005a, Bellazzini2008}. 

\citet{Siegel2007}, using HST photometry, suggested that at least four discrete stellar populations are present in the nucleus of the Sgr dSph galaxy, revealing a very complex star formation history.
These populations are:\\
\mbox{(i) [Fe/H]$=-1.8$}, \mbox{$[\alpha/\mathrm{Fe}]=+0.2$} with ages of 13\,Gyr;\\ 
\mbox{(ii) [Fe/H]$=-0.6$}, \mbox{$[\alpha/\mathrm{Fe}]=-0.2$} with ages of 4 to 6\,Gyr;\\
(iii) [Fe/H]$=-0.1$, $[\alpha/\mathrm{Fe}]=-0.2$ with ages of 2.3\,Gyr; and\\
\mbox{(iv) [Fe/H]$=+0.6$}, \mbox{$[\alpha/\mathrm{Fe}]=0.0$} with ages of 0.1 to 0.8\,Gyr.

Previous spectroscopic studies have focused on the division of the cluster into metal-rich and metal-poor subpopulations (i.e.~population (i) above vs. all others). \citet{Bellazzini2008} used $\sim$400 stars to show that the systemic velocities of the stars in both metallicity regimes coincide within $\simeq\pm1.0$\,km~s$^{-1}$, consistent with previous findings based on samples with a considerably lower number of stars \citep{Dacosta1995, Ibata1997, Monaco2005b}.
Despite this coincidence in radial velocities, the velocity dispersion profiles differ significantly for these populations, with the metal-rich stars having a much flatter dispersion profile than the metal-poor stars \citep{Bellazzini2008}.  
To explain the spatial coincidence and the differing dispersion and surface brightness profiles of the different populations, \citet{Monaco2005a} and \citet{Bellazzini2008} suggested the following possible scenarios:
(1) Both were formed {\em in situ} at the bottom of the potential well of the galaxy, first the metal-poor stars, followed by the metal-rich stars in subsequent star-forming episodes from enriched gas. 
(2) The metal-poor stars were an ordinary globular cluster that was driven by dynamical friction to the center of the galaxy where a nucleus had either already formed independently, or formed subsequently.

Using $N$-body simulations \citet{Bellazzini2008} showed that the latter scenario is feasible even with a pre-existing NSC. More specifically, the dynamical friction inspiral time scale of the metal-poor progenitor GC is $<$3~Gyr for a wide range of starting radii and initial relative velocities (unlike the other GCs in the Sgr dSph).  Furthermore, the end state of these simulations results in radial velocity differences between the pre-existing NSC and migrated GC of $<$2~km~s$^{-1}$.  

Studies on the abundances of the metal-poor and metal-rich population and radial variations of the metallicity have also been published. The metal-poor population displays a large spread in the iron content of its stars \citep[$\sigma_{\mathrm{[Fe/H]}}$$=0.186$,][]{Carretta2010b, Willman-Strader2012}. 
In addition, \citet{Carretta2010b} reported that a Sodium-Oxygen anticorrelation similar to those seen in other globular clusters is present in the metal-poor population. On the other hand, the authors do not observe any signatures of this anti-correlation in the metal-rich population.
\citet{Mucciarelli2017} found a metallicity gradient in the metal-rich population, with this component becoming more metal-poor at larger radii.  They also 
estimated that the chemical distribution they found for the Sgr dSph galaxy is consistent with a progenitor of M$_{DM}=6\times10^{10}$\,M$_\odot$, in agreement with the estimate by \citet{Gibbons2017}.

Finally, kinematic observations of the metal-poor population have also been used to determine if it hosts a central massive black hole.  Two studies have reported a possible detection based on dynamical modeling ($M_{BH} \sim 10{^4}$\,M$_{\odot}$, \citealt{Ibata2009, Baumgardt2017}).

All of these characteristics make the nucleus of the Sgr dSph galaxy a key object for understanding NSCs. Its proximity enables us to estimate parameters for individual stars, providing valuable information of the nucleus itself and its co-evolution with the progenitor galaxy. This object offers the opportunity to better understand the link between the most massive GCs and NSCs as stripped nuclei from low-mass galaxies. The MUSE spectrograph is an excellent instrument for studying individual stars in GCs  \citep[e.g.][]{Husser2016, Kamann2018}. In this work, we present a powerful MUSE data set centered on M54. With these data we perform a spectroscopic study of over 6\,500 stars, an order of magnitude increase over previous spectroscopic samples.

In this paper we present our data in Section\,\ref{data}, including the extraction of the single stellar spectra.  Section\,\ref{analysis} presents our analysis, including radial velocity, membership probability, metallicity and age measurements. Our results are presented in Section\,\ref{results}, we discuss these in Section\,\ref{discussion} and conclude in Section\,\ref{conclu}.

\section{Data} \label{data}

\subsection{Observations and data reduction}

The data set was acquired with MUSE \citep{Bacon2014}, an integral field spectrograph mounted at the UT4 of the Very Large Telescope at the Paranal Observatory in Chile, between June 29th and September 19th 2015 in run 095.B-0585(A) (PI: L\"utzgendorf).

A total of 16 pointings, with a field of view (FoV) of $59\farcs9 \times 60\farcs0$ each, create a 4$\times$4 mosaic that covers a contiguous area out to $\sim$2.5 times the half-light radius of M54 \citep[R$_{\mathrm{HL}}=0\farcm82$,][2010 edition]{Harris1996}, equivalent to $\sim$25 times its core radius \citep[R$_{\mathrm{c}}=0\farcm09$,][2010 edition]{Harris1996}. The estimated tidal radius of M54 is R$_{\mathrm{t}}=9\farcm87$ \citep[assuming a King-model central concentration of c$=2.04$,][2010 edition]{Harris1996}.

\begin{figure*}
\centering
\includegraphics[width=450px]{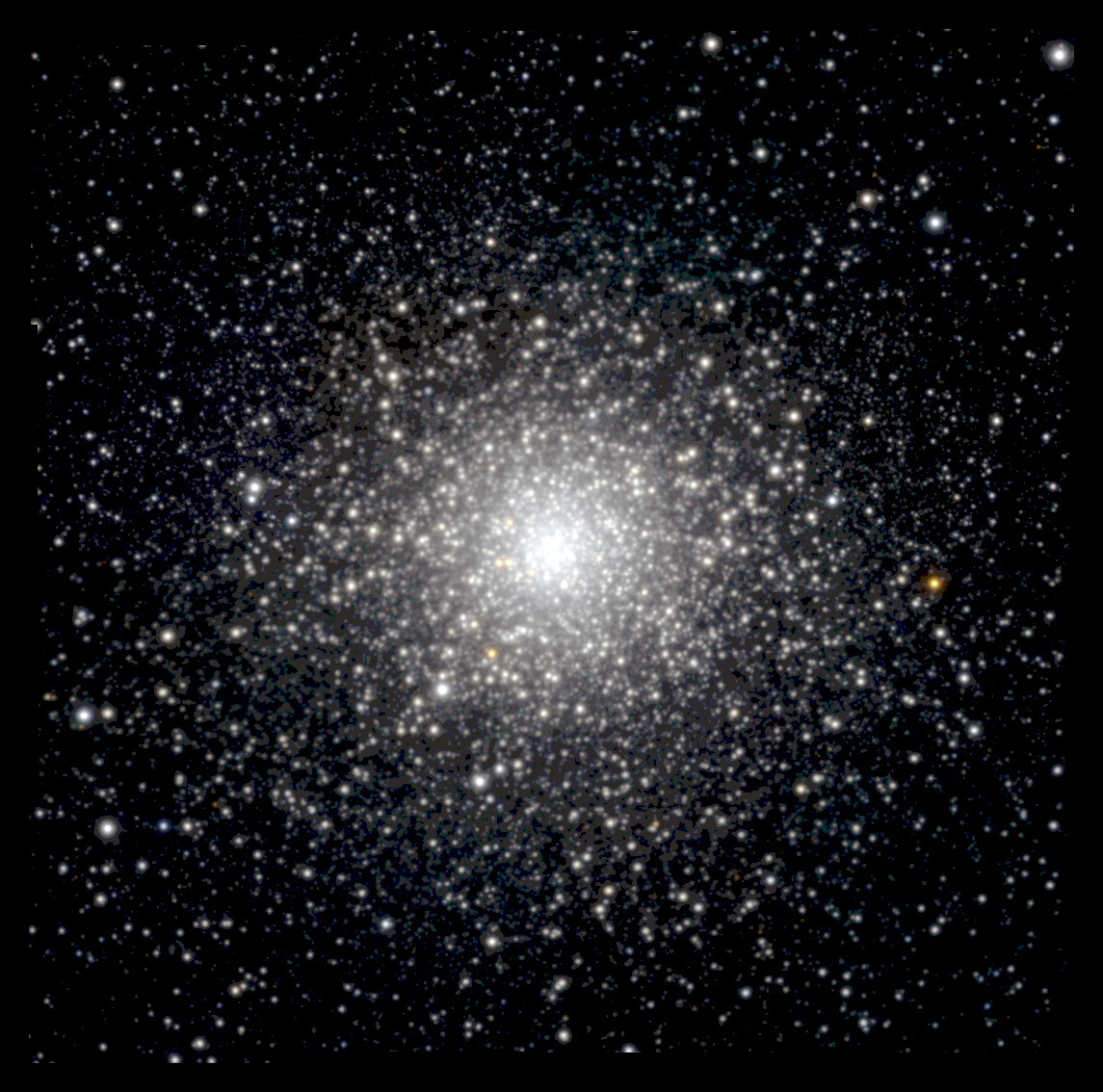}
\caption{Color image obtained from the MUSE data of Sgr dSph NSC, using synthetic i, r, and z filters. The image is a 4$\times$4 mosaic of 16 MUSE pointings, covering  $\sim$2.5\,R$_\mathrm{{HL}}$ (R$_{\mathrm{HL}}=0\farcm82$, \citealt[][2010 edition]{Harris1996}) of this NSC. The total covered field of view - considering overlaps - is $\sim 3\farcm5 \times 3\farcm5$, corresponding to 29\,pc $\times$29\,pc ($1\farcm=8.27$\,pc). North is up and East is to the left.}
\label{muse_mosaic}
\end{figure*}

Due to overlaps between neighboring pointings, the observations cover a field of view of $\sim$3\farcm5$\times$3\farcm5 ($1\farcm$ corresponds to 8.27\,pc at an assumed distance of $28.4$\,kpc, \citealt{Siegel2011}). This results in a total field of view of \mbox{29\,pc$\times$29\,pc}.
The data have a wavelength coverage of 4800\,-\,9300\,\AA~and a spectral resolution increasing with wavelength from \mbox{R$\sim1750$} to \mbox{R$\sim3750$}. The spatial sampling is $0\farcs2$/pix.
Each field was observed with three exposures, applying offsets in rotation of 90$\degree$ between them (no dithering) to increase the homogeneity of the data across the field of view. The exposure time for each of the three exposures was chosen to be 400\,s for the inner 4 pointings and 610\,s for the outer 12 pointings. This was done to avoid saturation of the innermost crowded region with high surface brightness.
During the observations, the airmass varied from 1.0 to 1.9; and the seeing between $0\farcs5$\,-\,$1\farcs2$. Figure\,\ref{muse_mosaic} shows the color image of the 4$\times$4 mosaic obtained from the MUSE data using synthetic i, r, and z filters.

The data reduction was performed using the MUSE pipeline \citep[Version 1.2.1,][]{Weilbacher2014}. The pipeline includes all tasks for the data reduction process: bias creation/subtraction, flat fielding, illumination correction and wavelength calibration. Before combining the single exposures, they were flux calibrated using a standard star and corrected for the barycentric motion of the Earth.

\subsection{Stellar spectra extraction} \label{stellar_extraction}

We extract individual stellar spectra with PampelMuse \citep{Kamann2013}. This tool takes a photometric catalogue as a reference for the positions and magnitudes of the stars in the field of view. It models the change in the point spread function (PSF) as a function of wavelength and allows deblending of sources efficiently even in crowded and dense regions.
A reference stellar catalogue for M54 was published by \citet{Siegel2007} as part of the ACS Survey of Galactic Globular Clusters \citep{Sarajedini2007}. The catalogue includes source positions and F606W (V) and F814W (I) magnitudes with their respective uncertainties. Its total coverage is larger than the MUSE data set FoV.

We extract the spectra of $\sim$55\,000 stars from the entire field. For our subsequent analysis we only consider the spectra with signal-to-noise ratio S$/$N$\geq$10, which are labeled by PampelMuse as good extracted spectra.
Within the entire MUSE FoV, we extract good spectra for a total of $\sim$7\,000 different stars in the magnitude range of $22<\mathrm{F814W}<13$  (I band). 

In our stellar sample we do not include main sequence (MS) stars below the oldest turn-off point since they are fainter than $\mathrm{I}\sim21$ and their spectra are typically below our $\mathrm{S/N}=10$ threshold.

\section{Analysis} \label{analysis}

To measure physical stellar parameters we use ULySS \citep[University of Lyon Spectroscopic Analysis Software,][]{Koleva2009}, a tool for determination of atmospheric parameters of stars through interpolating and fitting stellar library templates to the observed spectrum in the wavelength range of 3900 to 6800\AA. We note that the spectral wavelength range of MUSE allow us to fit from 4800 to 6800\AA\, only and we masked the Na I D line region between 5850 and 5950\,\AA\, of the observed spectrum. 
The observed spectrum is compared with a library of synthetic spectra with varying metallicity, surface gravity, and temperature. The best fit is achieved by interpolation within the synthetic library and $\chi^2$ minimization. The fit performed by UlySS is a Levenberg-Marquart local minimization for non-linear parameters, while for the linear ones uses a bounded-values least square. 
The synthetic spectroscopic grid available in ULySS is built on the basis of the ELODIE 3.2 library \citep{elodie}. It is limited to a metallicity range of $-2.5<$[Fe/H]$<0.5$.
ULySS also simultaneously estimates the radial velocity shifts of the input spectra. 
Finally, ULySS derives the uncertainties of the different physical parameters via Monte-Carlo simulations.
For stars in overlapping regions, where multiple spectra are extracted, the final values for the radial velocity and metallicity are the weighted mean and its uncertainty of the multiple ULySS measurements.

\begin{figure*}
\centering
\includegraphics[width=400px]{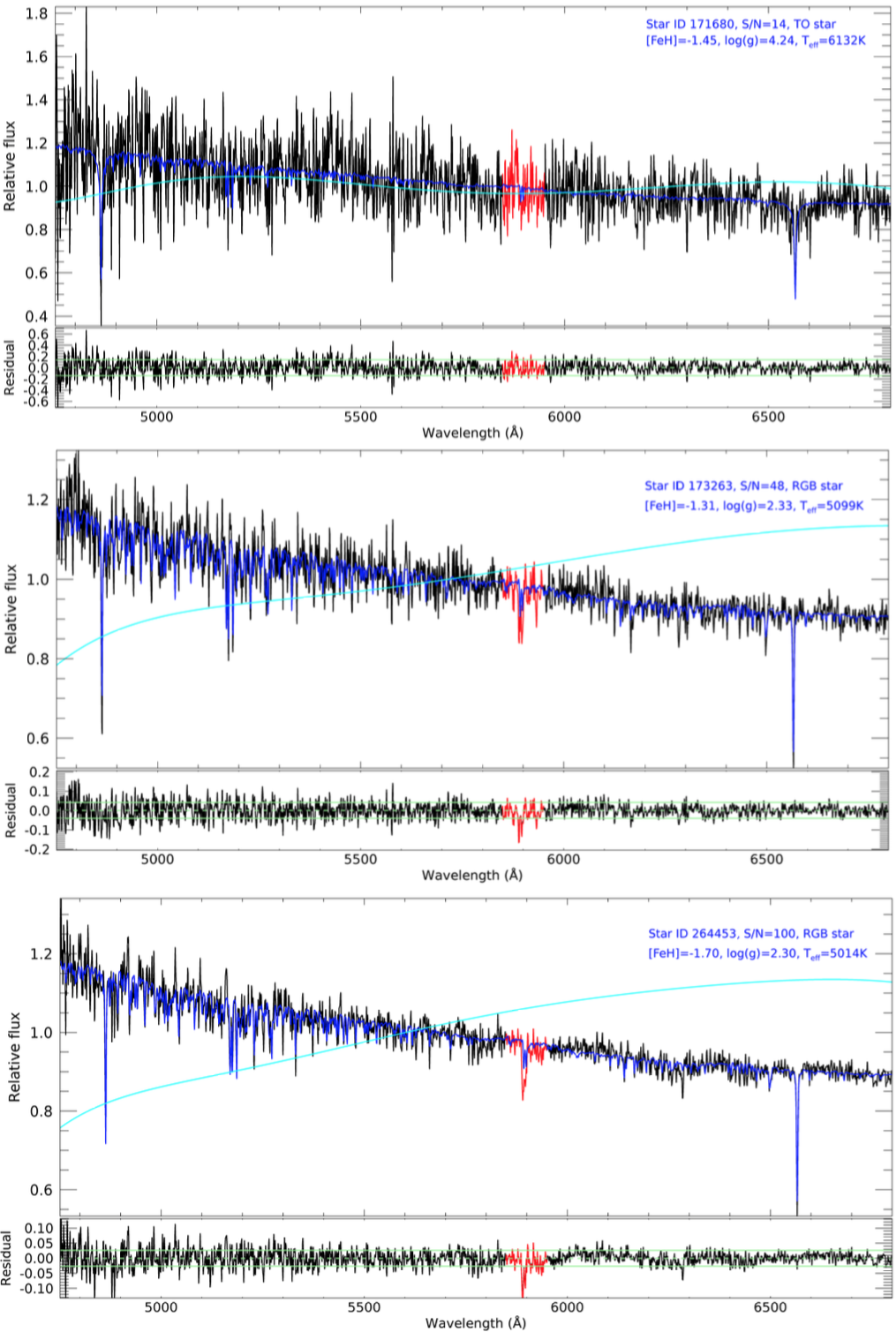}
\caption{Three examples of the fitting performed by ULySS  \citep{Koleva2009} for different types of stars with signal to noise of 14 (close to minimum of 10), 48 and 100, in the wavelength range of 4701 to 6800\AA\,.
Top panels: The best fit is represented in blue;  data not considered for the fit in red (both panels); and the multiplicative polynomial in turquoise. Bottom panels: Residual of the spectrum fitting (black). Overplotted in dashed and solid green are the mean and the 1$\sigma$ deviation, respectively.}
\label{ulyss}
\end{figure*}

Figure\,\ref{ulyss} shows three examples of the fitting performed by ULySS for stars with a spectrum of signal to noise of 14 (close to the minimum value of 10), 48 and 100. We use a multiplicative polynomial of order 6, which provides sufficiently good normalization of the observed spectra without the danger of overfitting.

\subsection{Radial velocity} \label{velocity_measurement}

From the spectral fitting performed by ULySS we obtain the radial velocities for each star. 

Since the space motion of the nucleus can mimic the effect of an additional rotation component, we correct the radial velocity values for this effect, known as perspective rotation, using equation 6 in \citet{vandeVen2006}.
We use the absolute proper motion of the cluster M54 \mbox{V$_W=-2.82\pm0.11$\,mas/yr} and \mbox{V$_N=-1.51\pm0.14$\,mas/yr} \citep{Sohn2015}. These values are in good agreement with the recent values based on \textit{Gaia} data by \citet{Vasiliev2019}. We obtain correction values between $-0.3$ and $0.3$\,km~s$^{-1}$.

In Appendix\,\ref{repeated_stars}, we discuss the agreement between velocity measurements derived for stars observed in multiple MUSE pointings.

\subsection{Sgr dSph NSC membership} \label{member_estimate} 

The position of the Sgr dSph galaxy with respect to the Milky Way and the large field of view of the MUSE data mosaic result in a considerable amount of contamination of non-member stars in our Sgr dSph NSC sample.
We determine the membership probability based on the iterative expectation maximization technique using "clumPy"\footnote{\url{https://github.com/bkimmig/clumpy}} \citep{Kimmig2015}. We use our radial velocity measurements and the distance of the stars to the center of the cluster M54 as input values (no metallicity estimates considered).

This procedure is described in detail by \citet{Walker2009} and later with some variations by \citet{Kimmig2015}. The technique iteratively estimates the mean velocity and velocity dispersion of the cluster plus the membership probability of each star until all the parameters converge. We use a foreground/background contamination model for the Milky Way based on the Besan\c{c}on model of the Galaxy \citep{Robin2003}.

\begin{figure}
\includegraphics[width=250px]{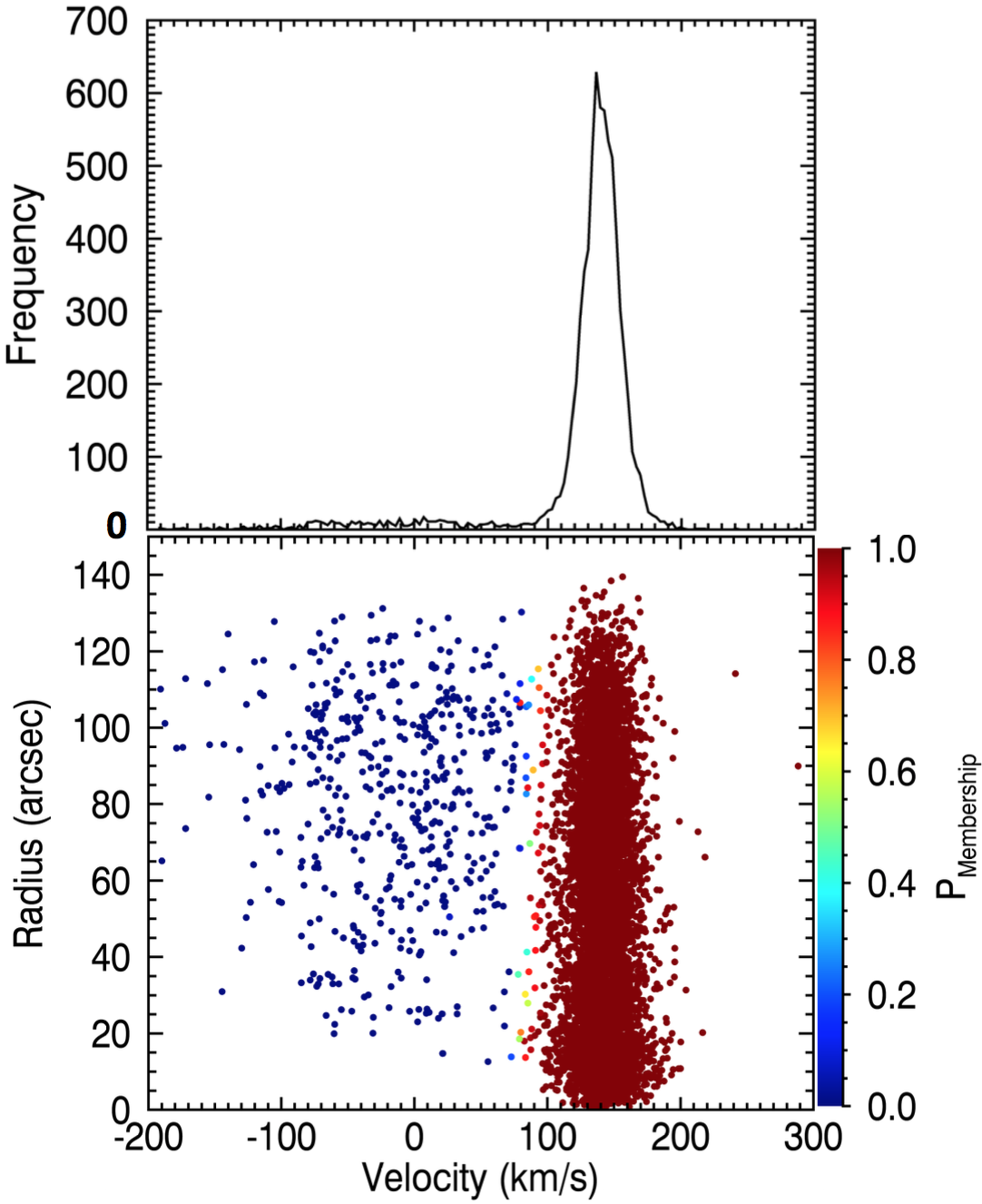}
\caption{{\bf Top:} Distribution of the radial velocities for the extracted stars with S/N$\geq10$ from the entire MUSE FoV. {\bf Bottom:} Radius versus velocity for our stellar sample color coded by the cluster membership probability (P$_{\mathrm{Membership}}$). We consider as member stars of the Sgr dSph NSC those with probability $\geq70\%$.}
\label{membership}
\end{figure}

The membership probability obtained with the expectation maximization technique for each star is represented by color in the radius versus velocity plot in Figure\,\ref{membership}.
From our total sample of 7\,000 stars we consider those with probability \mbox{$\geq70\%$} to be members of the Sgr dSph NSC, leaving a total of 6\,651 members. 

We obtain a maximum likelihood median radial velocity for the member stars (dark red points) of $141.34\pm0.18$\,km~s$^{-1}$ and a median central velocity dispersion of $16.31\pm0.28$\,km~s$^{-1}$. The radial velocity estimate is in good agreement with previous works, e.g V$_r\sim141$\,km~s$^{-1}$ by \citet{Bellazzini2008} for the metal-poor and metal-rich populations.
From this point, the analysis is carried out on a sample that only includes the members of the Sgr dSph NSC determined by this technique.

\begin{figure}
\includegraphics[width=250px]{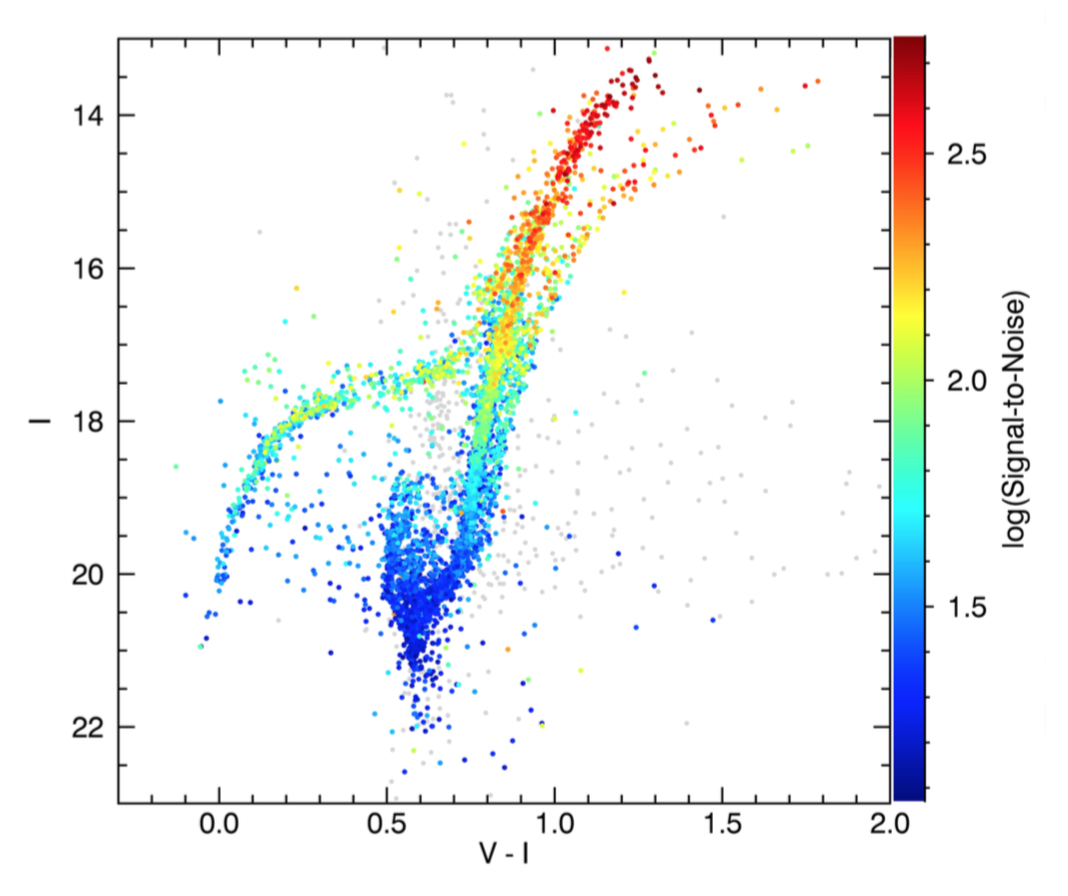}
\caption{CMD for the total number of extracted spectra with \mbox{S/N$\geq10$}. Stellar photometry information in F606W (V) and F814W (I) filters (HST/ACS WFC) taken from the M54 catalogue \citep{Siegel2007}  of the ACS Survey of Galactic Globular Clusters \citep{Sarajedini2007}. Member stars of the Sgr dSph NSC (probability $\geq70\%$) are color coded by the signal-to-noise logarithm  and non-members  (probability $<70\%$) are shown in gray.}
\label{cmd_sn}
\end{figure}

In the color-magnitude diagram (CMD) in Figure\,\ref{cmd_sn} we show the full sample of extracted single stellar spectra. Member stars of the Sgr dSph NSC (probability $\geq70\%$) are color coded by the signal-to-noise logarithm, while non-members (probability $<70\%$) are shown in gray. The stellar photometry information in F606W (V) and F814W (I) filters are extracted from the M54 HST reference catalogue.

\subsection{Metallicity from full spectrum fitting}  \label{met_ulyss}

\begin{figure}
\centering
\includegraphics[width=230px]{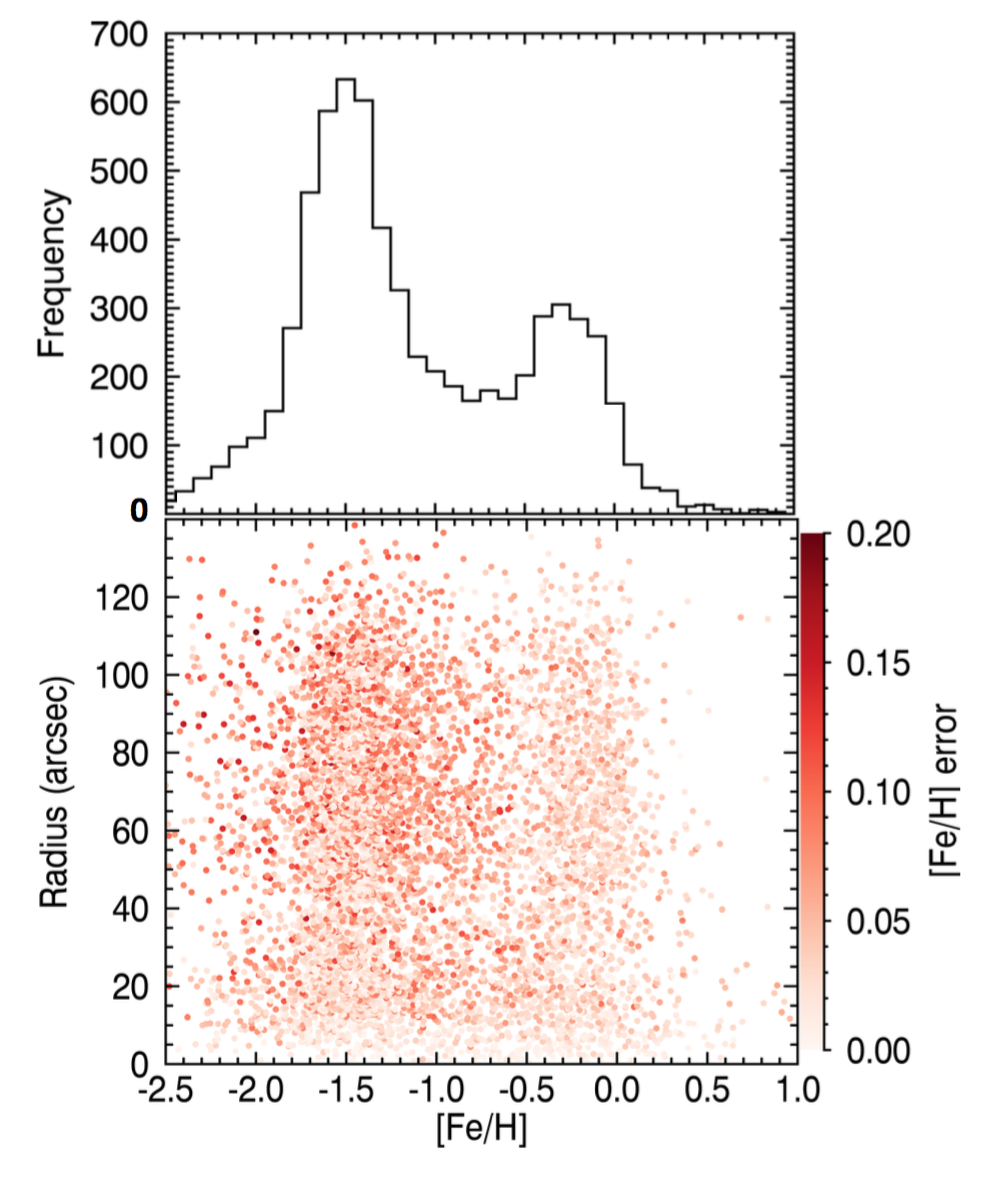}
\caption{\textbf{Top:} Metallicity histogram of the member stars of Sgr dSph NSC. \textbf{Bottom:} Radius versus metallicity plot for the same sample. The stars are color coded by the uncertainties in the [Fe/H] measurements.}
\label{rad_met}
\end{figure}

The top panel of Figure\,\ref{rad_met} shows the metallicity histogram for the member stars of the Sgr dSph NSC. This metallicity distribution is highly consistent with the one presented by \citet{Mucciarelli2017} for a total of 76 stars in a radius range of $0\farcm0<\mathrm{R}\leqslant2\farcm5$. We cross matched our sample with those from \citet{Carretta2010a} and \citet{Mucciarelli2017}, and compare the measurements for the stars we have in common. The values are consistent within the errors. We obtain a mean difference of 0.15$\pm$0.03\,dex, when comparing with \citet{Carretta2010a}, with a standard deviation of $\sigma=0.16$. From the comparison with \citet{Mucciarelli2017}, we obtain a mean difference of 0.05$\pm$0.02\,dex with a standard deviation of $\sigma=0.17$. This comparison is included in Appendix\,\ref{feh_comparison}.

In the bottom panel, we present the relation between radius and metallicity for the sample of 6\,651 stars as derived with ULySS, color coded by the metallicity uncertainties. Two clear distributions in metallicity become apparent: a metal-poor population at [Fe/H]$=-1.5$ and a metal-rich population around [Fe/H]$=-0.3$ with a separation at [Fe/H]$=-0.8$. The median metallicity uncertainty is 0.05\,dex.

The top panels in Figure\,\ref{cmds} show the CMD color coded by metallicity (hereafter CMD+metallicity) corresponding to all the Sgr dSph NSC members (and a zoom-in onto the turn-off region, top right panel). We overplotted in blue, green and orange isochrones from the Dartmouth Stellar Evolution Database \citep{Dotter2008} corresponding to different stellar populations in the Sgr dSph NSC, using as a reference the work by \citet{Siegel2007}. 
We are able to see stars in different evolutionary stages: turn-off point (TO), red giant branch (RGB), horizontal branch (HB), asymptotic giant branch (AGB), red clump (RC), even the extreme horizontal branch (EHB) and the blue plume (BP). The CMD shows clear evidence for an old metal poor population, as well as at least two distinct younger populations with metallicity $\gtrsim-0.4$ that agree well with the 4.5-6\,Gyr isochrone highlighted by \citet{Siegel2007}.

For our complete sample of member stars we estimate a mean metallicity of [Fe/H]$=-1.06$, with a standard deviation of $\sigma=0.64$.        
For the metal poor stars (\mbox{[Fe/H]$<-0.8$}) we obtain a mean metallicity of  \mbox{[Fe/H]$=-1.45$} and a standard deviation of $\sigma=0.32$. This is in good agreement with previous works, such as \citet{Brown1999} and \citet{Dacosta1995} with an estimate of [Fe/H]$=-1.55$, and later by  \citet{Carretta2010a}, who estimated a metallicity of \mbox{[Fe/H]$=-1.559\pm0.021$\,dex}.

In the case of the stars in the metal rich regime (\mbox{[Fe/H]$>-0.8$}), we obtain a mean value of [Fe/H]$=-0.27$ and a standard deviation of $\sigma=0.29$. In this case, the metallicity measurement differs from previous studies. \citet{Carretta2010a} obtained a value of \mbox{[Fe/H]$=-0.622\pm0.068$\,dex}, and \citet{Mucciarelli2017} \mbox{[Fe/H]$=-0.52\pm0.02$\,dex}. A possible explanation for this difference is discussed in Section\,\ref{3pop}.

We repeated the metallicity measurements with a second method, using Ca\,II triplet lines, and obtained consistent metallicity measurements using the two independent methods (see Appendix \ref{caii}). 

\subsection{Stellar age estimates} \label{stellar_age_estimates}
 
Valuable information about the star formation history (SFH) of the Sgr dSph NSC can be obtained from age and metallicity estimates together.
For the age estimates of individual stars we use a grid of isochrones, which we compare with the magnitudes and colors of the stars as measured from the HST photometry in V (F606W) and I (F814W) bands with their respective uncertainties ($\sigma_\mathrm{V}$, $\sigma_\mathrm{I}$), and metallicities Z, obtained from the MUSE spectra, including their uncertainties $\sigma_{\mathrm Z}$. We set the metallicity uncertainty to 0.1\,dex when lower than that value.
We consider age the only free parameter of the model.

We use Bayes' theorem to estimate the ages through the posterior probability as: 
\begin{equation}
\mathrm{ P(\tau\,|\,V_{obs},I_{obs},Z_{obs})\,\propto\,P(V_{obs},I_{obs},Z_{obs}\,|\,\tau)}\times \mathrm{P}(\tau),
\end{equation}
where the normalized likelihood function P(V$_{\mathrm{obs}}$,I$_{\mathrm{obs}}$,Z$_{\mathrm{obs}}$\,|\,$\tau$) of the observables at a given age is a trivariate Gaussian:

\begin{equation}
\begin{split}
P(\mathrm{V_{obs},I_{obs},Z_{obs}\,|\,\tau})\,=\, & \frac{1}{\mathrm{ \sigma_{V_{obs}} \, \sigma_{I_{obs}} \, \sigma_{Z_{obs}}  \, (2\pi)^{3/2}}}   \\
         & \times \mathrm{exp\frac{ -(V_{obs} - V_{0})^2 }{2\sigma_{V_{obs}}^2}}  \\
         & \times \mathrm{exp\frac{ -(I_{obs} - I_{0})^2}{2 \sigma_{I_{obs}}^2}} \\
         & \times \mathrm{exp\frac{ -(Z_{obs} - Z_{0})^2}{2 \sigma_{Z_{obs}}^2} } ,
\end{split}
\label{lkhd}
\end{equation}
where V$_0$, I$_0$, and Z$_0$ denote the real magnitudes and metallicity of the star given its age and the $\sigma$ values represent the independent errors in each measurement .
We marginalize over the unknown true magnitudes and metallicity of the star by integrating over the isochrones \citep[see e.g.][for a similar approach]{Pont2004}.

Due to the lack of Extended Horizontal Branch isochrone models, we have excluded stars with colors between \mbox{${\mathrm V-\mathrm I}=-0.2$} and 0.4 and I in the range of 21 to 17 in our age analysis.

Initially, we used a flat prior in age P($\tau$) for all the remaining stars. The age-metallicity relation (AMR) found for the TO region stars turned out to be very well represented by the empirical model published by \citet{Layden_Sarajedini2000}. On the other hand, the obtained ages for a fraction of metal-poor RGB stars were lower than expected for such metallicity regime. This degeneracy could be a consequence of high abundance of elements like oxygen, which has been reported to display a high spread in this metal-poor population \citep{Carretta2010a}. In fact, \citet{vandenberg2012} showed the impact of different elemental abundances on computed evolutionary tracks at similar [Fe/H] values and how this affects the color of the stars, thus their position in the CMD. After this examination of the initial ages, to guard against systematic age uncertainties, we set a weak Gaussian age prior P($\tau$) to their model with a standard deviation of $3$\,Gyr for the age estimates of all extracted member stars from the MUSE cube.

We built the isochrone grid with a resolution of 0.2\,Gyr with ages from 0.2 to 15\,Gyr from the Dartmouth Stellar Evolution Database \citep{Dotter2008} assuming the HST/ACS WFC photometric system. The metallicity range is from $-2.495$ to $0.500$\,dex with steps of $0.005$\,dex, considering $[\alpha/\mathrm{Fe}]=0.0$. 
We correct the isochrone magnitudes by extinction of A${_ \mathrm{V}}=0.377$ and A${_\mathrm{I}}=0.233$ from NED\footnote{The NASA/IPAC Extragalactic Database (NED) is operated by the Jet Propulsion Laboratory, California Institute of Technology, under contract with the National Aeronautics and Space Administration.}. These are obtained using the calibration published by \citet{Schlafly2011} for the F606W and F814W bands for a reddening of $E(B-V)=0.15$ estimated in the central $5\farcm$ radius of M54 \citep{SFD1998} with  R$_\mathrm{V}=3.1$. We shifted the grid of isochrones adopting a distance modulus of $(m-M_0)=17.27$ \citep{Siegel2007}. For comparison, we performed the same analysis considering a reddening of $E(B-V)=0.16$ and a distance modulus of $(m-M_0)=17.13$ \citep{Sollima2010}, obtaining consistent age estimates.
We take the mode of the likelihood distribution of ages as the best age estimate and the highest density interval that accommodates 68\% of the probability as the age uncertainty. 

\begin{figure*}
\centering
\includegraphics[width=500px]{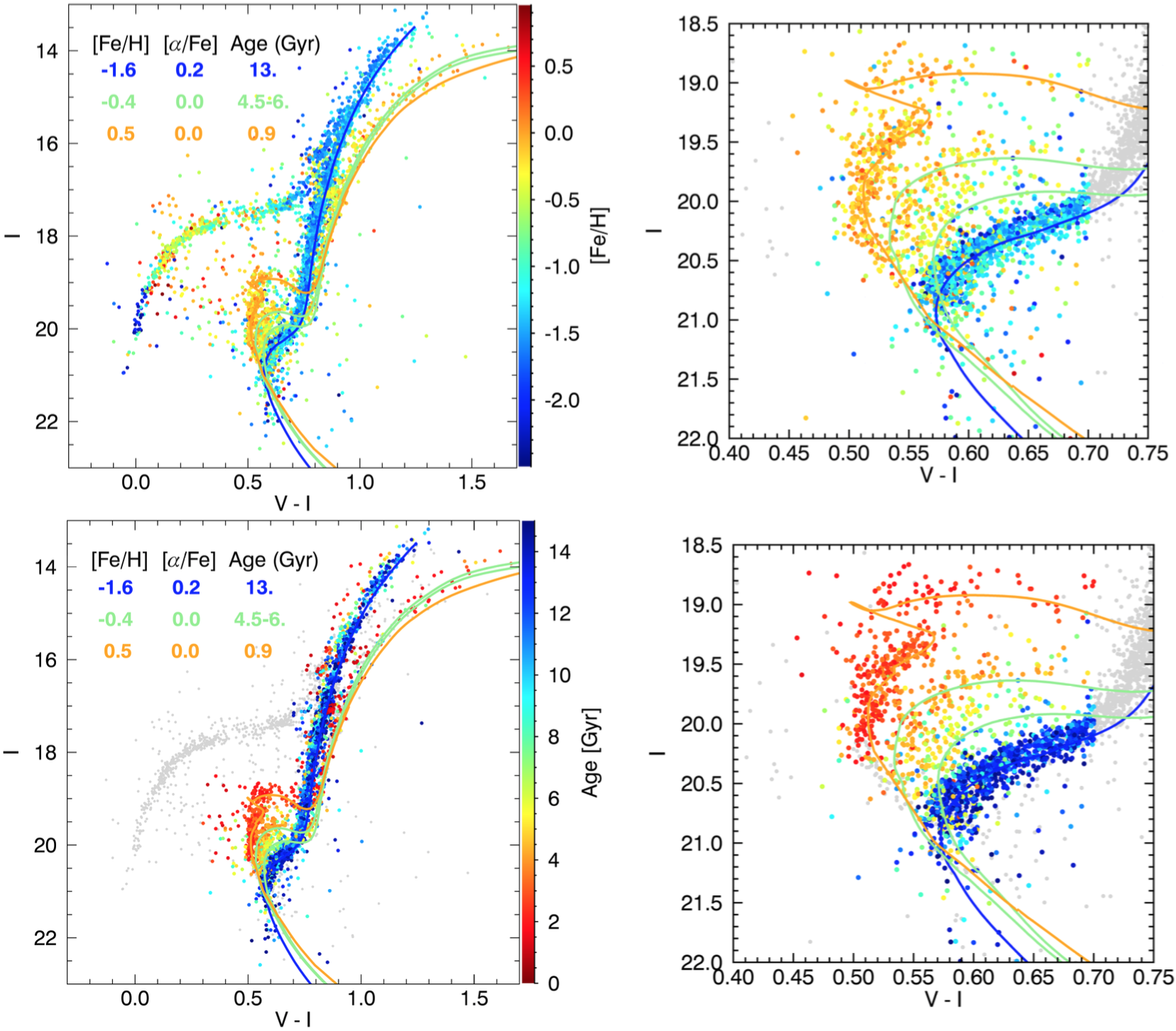}
\caption{Left panels: CMDs for the Sgr dSph NSC members color-coded by metallicity (CMD+metallicity) estimated with ULySS at the top and color coded by age (CMD+age) at the bottom.
Right panels: Zoom into the TO region.
Updated isochrones from the Dartmouth Stellar Evolution Database are overplotted in blue, green and orange for three stellar populations published by \citet{Siegel2007}.}
\label{cmds}
\end{figure*}

The CMD for the clean stellar sample color coded by age (hereafter CMD+age) is presented in the bottom left panel of Figure\,\ref{cmds}, with a zoom-in to the TO region in the bottom right panel of the figure.
Three isochrones based on those presented in \citet{Siegel2007} are overplotted; they are also from the Dartmouth Stellar Evolution Database \citep{Dotter2008}.  

\subsection{Populations split by age and metallicity} \label{pop_split}

We use Gaussian Mixture Models (GMM) in age-metallicity space to separate the stars into distinct multiple populations.
To this aim, we used only TO region stars, where we obtain the most reliable age estimates as we describe in Section\,\ref{stellar_age_estimates}.
We tested GMM with three, four, and five components and found that a GMM with four components gives the best representation of our data set based on the Akaike Information Criterion \citep[AIC,][]{Akaike1974}.
We optimized the GMM using the Affine-Invariant Markov Chain Monte Carlo (MCMC) algorithm \citep{Goodman2010, Foreman-Mackey2013}.
We then computed the probabilities of all stars in the observed sample of belonging to any of the GMM defined populations. The fourth Gaussian component of the GMM captures outliers and is not further considered in our analysis.

We estimate the mean and spread in both age and metallicity for each of the subpopulations. We start by taking stars with a high probability ($>50$\%) of belonging to each subpopulation.  We then perform a maximum likelihood fit for the ages and metallicities of these stars using a 2-dimensional Gaussian and accounting for the observational uncertainties in each stars' age and metallicity. The result is a mean and {\em intrinsic} spread in both age and metallicity for each population. We assumed that the [Fe/H] Probability Distribution Functions (PDF) are Gaussian and integrated numerically over the non-Gaussian individual stellar age PDFs.
From the results of this procedure we define three different subpopulations as follows:

\begin{itemize}
\item YMR: young metal rich (red), with mean age of $2.16\pm0.03$\,Gyr and mean \mbox{[Fe/H]$=-0.04\pm0.01$}.
\item IMR: intermediate age metal rich (orange), with mean age of $4.28\pm0.09$\,Gyr and mean \mbox{[Fe/H]$=-0.29\pm0.01$}.
\item OMP: old metal poor (blue), with mean age of $12.16\pm0.05$\,Gyr and mean \mbox{[Fe/H]$=-1.41\pm0.01$}.
\end{itemize}

We find that the YMR population has an intrinsic $1\sigma$ spread of $0.20\pm0.03$\,Gyr in age and $0.12\pm0.01$\,dex in metallicity; the IMR population has an intrinsic spread of $1.16\pm0.07$\,Gyr in age and $0.16\pm0.01$\,dex in metallicity; the OMP population has an intrinsic spread of $0.92\pm0.04$\,Gyr in age and $0.24\pm0.01$\,dex in metallicity. The uncertainties are represented by the standard deviation in each parameter in the converged part of the MCMC, thus obtaining low errors. The age-metallicity correlation coefficients, $\rho$, for the three populations are $\rho_{\mathrm{YMR}}=-0.35$, $\rho_{\mathrm{IMR}}=-0.70$ and $\rho_{\mathrm{OMP}}=-0.97$.

To check for consistency, we use the same method to estimate the intrinsic spread in metallicity for stars with \mbox{$\mathrm{I}\leqslant16.5$\,mag} where the signal-to-noise of the stars is mostly $>100$. This leaves a total of 22 stars for the YMR (5\% of YMR sample) and 208 for the OMP ($\sim9$\% of OMP sample). For the YMR subpopulation we obtain an intrinsic $1\sigma$ spread of $0.11\pm0.06$\,dex, and $0.13\pm0.03$\,dex for the OMP subpopulation. The spread for the YMR subpopulation derived from the brightest stars is consistent with the one measured for the stars of all magnitudes. For the OMP subpopulation, we observe a difference of 0.11\,dex. This can be consequence of the low number of stars in comparison with the entire sample, and the effects of the age uncertainties at the RGB region, which can be more uncertain, as we mention in the previous section. Since these values are obtained from a reduced fraction of the entire sample ($<10$\%), we will consider the intrinsic spread measured from the sample at all magnitudes.

\begin{figure*}
\centering
\includegraphics[width=490px]{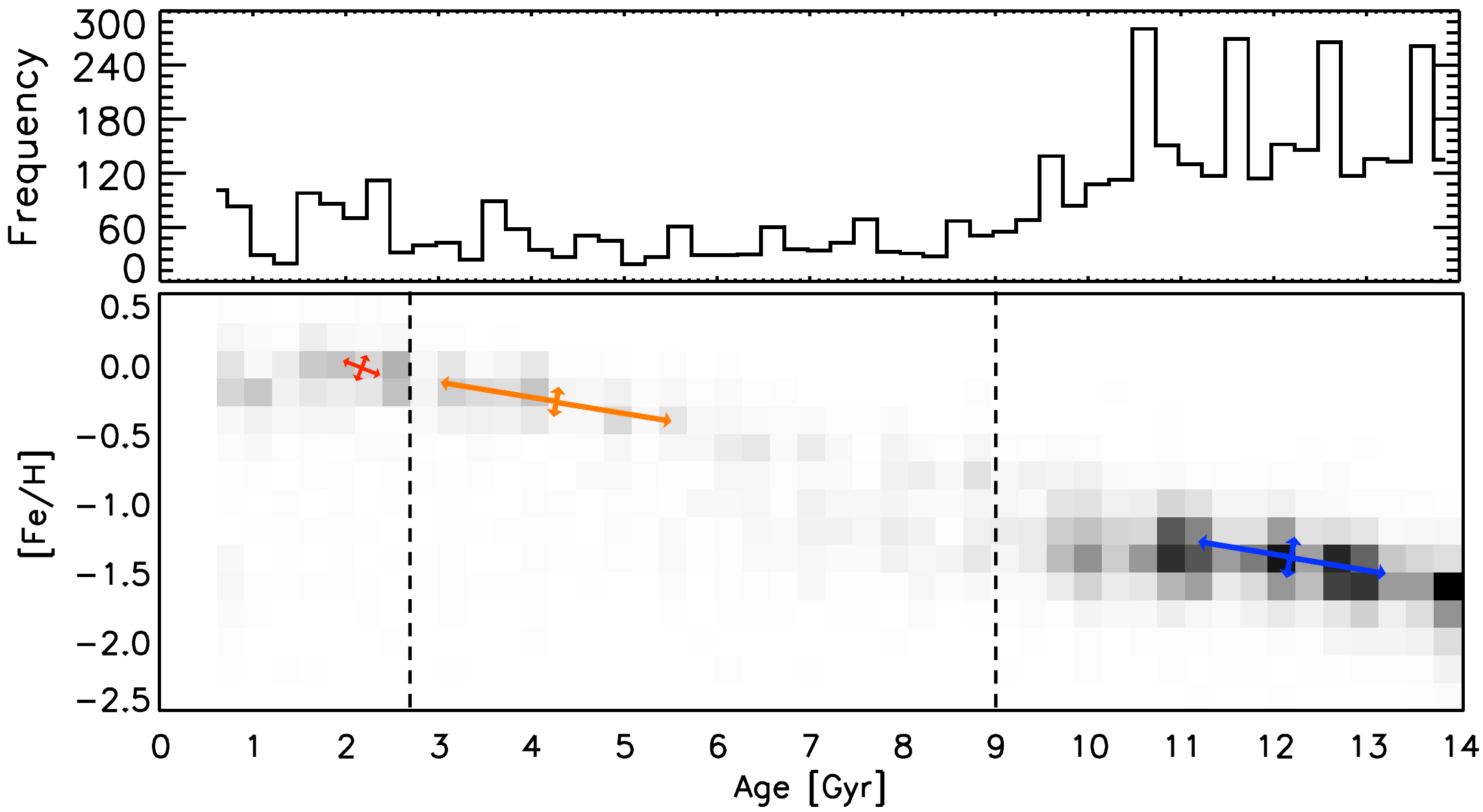}
\caption{\textbf{Top}: Age histogram. 
\textbf{Bottom}: Density plot of the age-metallicity relation.
The crosses show the intrinsic spread of the different stellar populations in the Sgr dSph NSC: young metal rich (YMR) in red, intermediate age metal rich (IMR) in orange and old metal poor (OMP) in blue. The subpopulations stellar parameters are summarized in Table\,\ref{table}. All the stars included in both panels have age relative errors $\leqslant$40\%. The vertical dashed lines show where the probability is equal for the two neighbouring subpopulations. {\bf Note:} The inclination of the crosses is given by the correlation coefficient obtained from the Gaussian mixture model.}
\label{amr}
\end{figure*}

The AMR is presented in Figure\,\ref{amr}, where we include all stars for which ages were measured and have relative age errors $\leq40\%$. The overplotted crosses represent the intrinsic spreads of the different populations: YMR in red, IMR in orange, and OMP in blue. 
The intrinsic age and metallicity spreads together with the age-metallicity correlation coefficient $\rho$ define the inclination angle of the crosses for each subpopulation.
We discuss the possible origin of these populations in Section \ref{discussion}.

For the RC stars in the sample we were able to obtain reliable measurements in metallicity and found that they all fall in the metal-rich regime. The complexity of this stage of stellar evolution prevents us from getting reliable ages for the RC stars from simple isochrone fits.
Due to this effect, the RC stars were not considered to belong to any of the subpopulations when we decompose the stars into subpopulations via the maximum likelihood method.
For completeness, these stars are shown in Figure\,\ref{amr} and fall on the diagram clustered together as the overdensity at $\sim$1\,Gyr in a metallicity range of $-0.5<\mathrm{[Fe/H]}<0.0$.
To get a better estimate of the likely age distribution of RC stars, we follow \citet{Girardi2016} and compute the RC age distribution for representative SFHs of Sagittarius. We do this for two assumptions of the SFH: (i) Sgr dSph galaxy has a constant SFH and (ii) SFH follows the observed SFH from the CMD analysis of \citet{Boer2015}. 
The \citet{Boer2015} SFH shows a declining SFR with time, but is constructed from the tail of the Sgr dSph, so is likely missing the most metal rich and young stars - biasing the star formation to early times.  Inclusion of the younger populations in the CMD analysis might produce a SFH more closely represented by a constant SFH.  We therefore consider these two cases as plausible bounds on the likely true SFH of Sgr. For the constant SFH case we find that the RC stars are predominantly ($\sim51$\%) younger than 2.2\,Gyr, thus being more likely to belong to the YMR subpopulation than to the IMR (10\%.) 
For case (ii), we obtain that anywhere from 10 to 70\% of RC stars are younger than 2.2\,Gyr, with probabilities of less than 10\% for the IMR.
Taking these two cases into consideration, we consider it most likely that the majority of the RC stars is associated with the YMR subpopulation, and consider them as part of that population in our further analysis.

\section{Results} \label{results}

\subsection{Population analysis}  \label{3pop}

Comparing the metallicity and age information presented in the CMDs in Figure\,\ref{cmds}, we observe that in the TO region in the CMD+age we distinguish three, well separated subpopulations with different age ranges. The faintest stars have ages over 9\,Gyrs. This subpopulation belongs to the metal poor regime in the CMD+metallicity. The middle sub-giant branch population has ages of $\sim$3-9\,Gyr, while the brightest stars have ages younger than 3\,Gyr. These last two populations belong to the metal rich regime in the CMD+metallicity. From the CMD+metallicity figure, some small difference in metallicity ($\sim$0.3 dex) is also visible between these two younger populations.

In the CMD+metallicity the metal poor RGB is well defined by stars with metallicity below \mbox{[Fe/H]$=-0.8$}, displaying a wide range toward \mbox{[Fe/H]$=-2.5$}, corresponding to an intrinsic iron spread of  \mbox{$\sigma_{\mathrm{[Fe/H]}}=0.24$\,dex}.
In addition, we observe that the stars in the blue plume region between $18.5\leqslant \mathrm{I} \leqslant20.5$ and $0.1\leqslant \mathrm{V-I} \leqslant0.5$ show a wide range in metallicity (from $-2.0$ to 0.5). We discuss this further in Section\,\ref{IMR}.

From the CMDs and the age-metallicity relation (Figure\,\ref{amr}), we can decipher the star  formation history of the Sgr dSph NSC. It appears to be extended from 0.5 to 14\,Gyr, showing a clear metallicity enrichment towards younger stars. 

\begin{figure}
\centering
\includegraphics[width=245px]{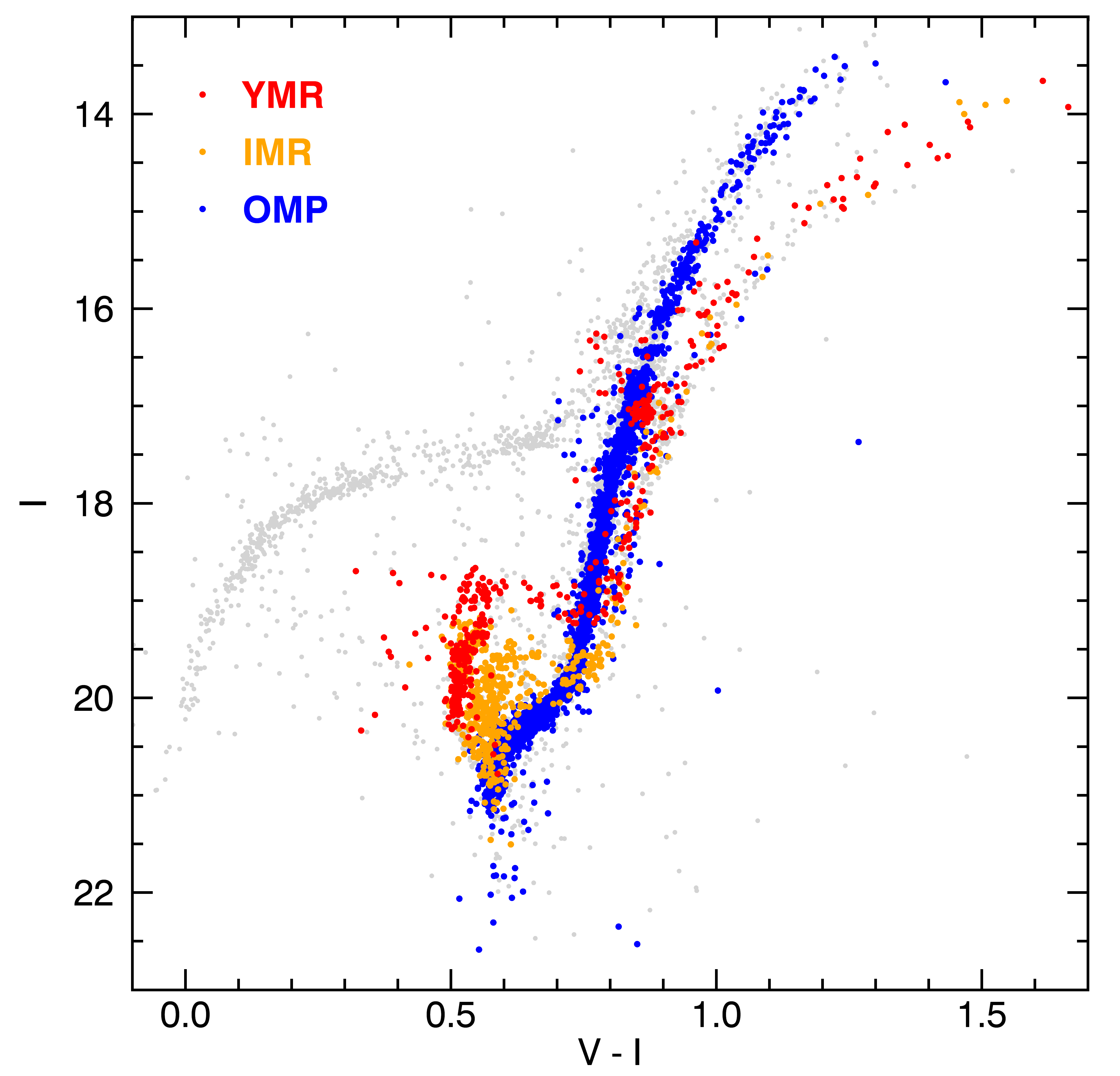}\\
\includegraphics[width=220px]{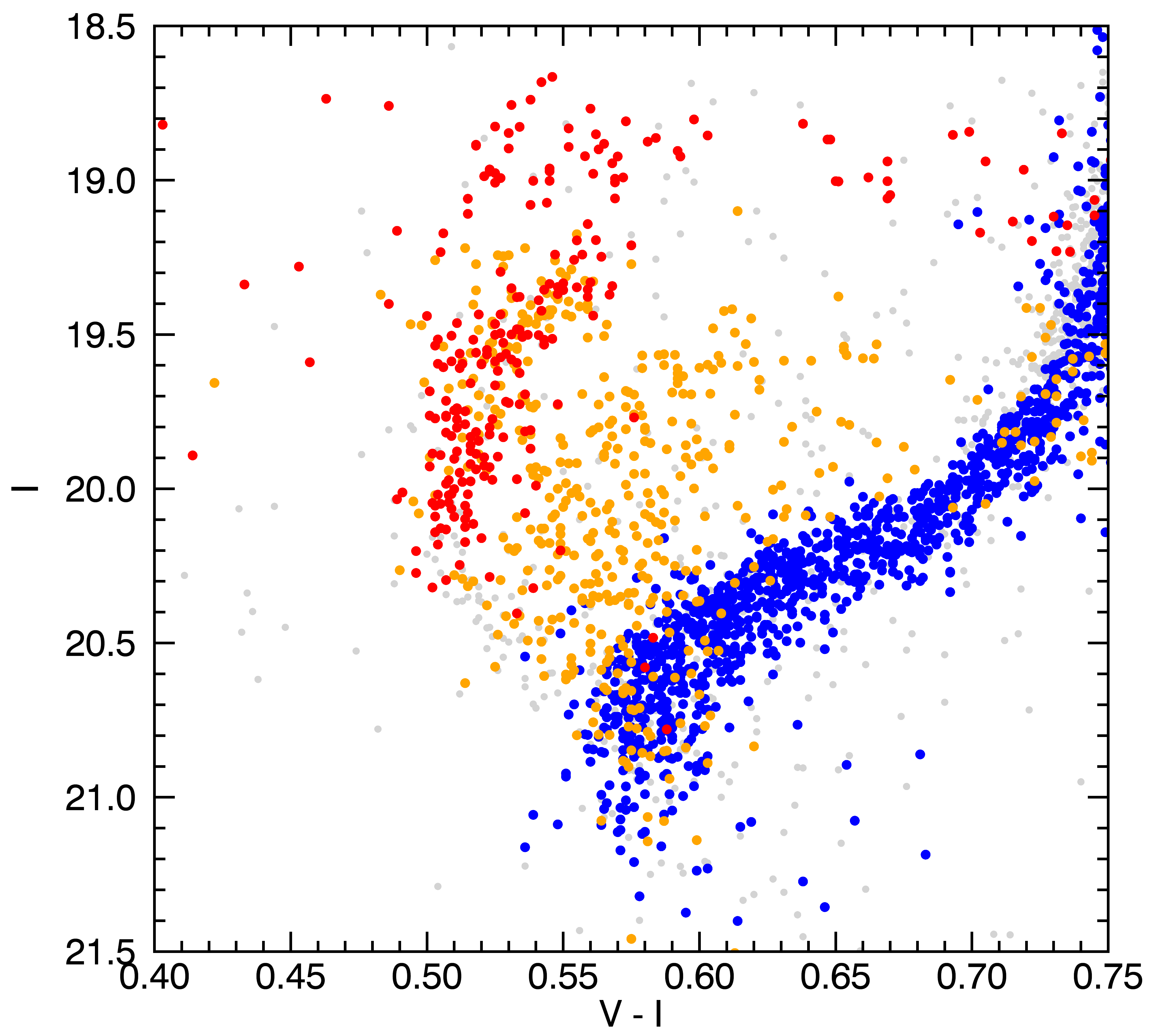}\\
\caption{For the two panels: red points represent the YMR subpopulation ($2.2$\,Gyr,  \mbox{[Fe/H]$=-0.04$}), orange points for the IMR (4.3\,Gyr, \mbox{[Fe/H]$=-0.29$}) and blue for the OMP ($12.2$\,Gyr, \mbox{[Fe/H]$-1.41$}). Gray points show the stars with age relative error greater than 40\% or for which age was not estimated. Top panel: Color-magnitude diagram.  Bottom panel: zoom into the top panel showing just the TO region.}
\label{3pop_cmd}
\end{figure}

We present a CMD comparing the three subpopulations in Figure\,\ref{3pop_cmd}.
In both panels the subpopulations are represented as: YMR in red, IMR in orange and OMP in blue. In the top panel we can see how the different populations fall in different positions in the CMD, this is easier to see in the zoom into the TO region showed in the bottom panel.

Our findings are in good agreement with the characteristics of three (of four) subpopulations obtained with isochrone fitting by \citet{Siegel2007}, and later also by \citet{Mucciarelli2017}. From their fitting of the CMD, \citet{Siegel2007} suggest the presence of four populations: (i) the old metal poor, with ages of 13\,Gyr and  [Fe/H]$=-1.8$, (ii) intermediate age metal rich, with ages between 4-6\,Gyr and [Fe/H]$=-0.6$, (iii) young metal rich population with 2.3\,Gyr and [Fe/H]$=-0.1$, and (iv) very young, very metal-rich population with [Fe/H]$=0.6$ and ages 0.1-0.8\,Gyr. We do not detect any stars belonging to their fourth population. 

We do find a higher mean metallicity for the metal-rich regime population compared to \citet{Carretta2010a} and \citet{Mucciarelli2017}.  This may be due to several factors.  First, we separate the metal-rich stars into two subpopulations, with the YMR having significantly higher mean metallicity than the IMR; the latter, older stars have metallicity closer to previous measurements. Second, our spatial coverage is confined to much smaller radii in comparison to \citet{Carretta2010a} and \citet{Mucciarelli2017}, which cover out to a larger radius. Third, the resolution of our data allow us to extract more stars from the most crowded central regions. Fourth, on average, our metallicity estimates are slightly higher. We obtain a mean $\Delta$[Fe/H] of 0.15$\pm$0.03\,dex when comparing with \citet{Carretta2010a}, and 0.05$\pm$0.02\,dex with \citet{Mucciarelli2017}. The differences in the measurements might be caused by the method used and/or spectral resolution (see Appendix\,\ref{feh_comparison}).

In Table\,\ref{table} we include a summary of all the estimated parameters for the subpopulations. 

\begin{deluxetable*}{c|ccc}
\tablecaption{Summary of the stellar subpopulations in M54.  \label{table}}
\tablehead{
\colhead{subpopulations}  &  \colhead{YMR} & \colhead{IMR} & \colhead{OMP} }
\startdata
  [Fe/H]                       & $-0.04\pm0.01$  & $-0.29\pm0.01$   & $-1.41\pm0.01$      \\
  $\sigma_{\mathrm{[Fe/H]}}$   & $0.12\pm0.01$    & $0.16\pm0.01$   & $0.24\pm0.01$ \\
  Age (Gyr)                    & $2.16\pm0.03$    & $4.28\pm0.09$   & $12.16\pm0.05$ \\ 
 $\sigma_{\mathrm{Age}}$ (Gyr) & $0.20\pm0.03$    & $1.16\pm0.07$   & $0.92\pm0.04$  \\ 
  Correlation factor ($\rho$)  & $-0.35$    & $-0.70$   & $-0.97$ \\ 
  Median V${_r}$ (km~s$^{-1}$) & $141.92\pm0.54$  & $142.61\pm0.59$   & $141.22\pm0.26$  \\
  RA Centroid (degrees)        & $283.76351\pm0.00176$  &  *  & $283.76299\pm0.00094$   \\
  Dec Centroid (degrees)       & $-30.476992\pm0.000977$     &  *  & $-30.477067\pm0.000673$   \\
  Half-light radius (arcmin)   & $1.47\pm0.20$    &  *  & $1.90\pm0.12$	\\
  Ellipticity                  & $0.31\pm0.10$    &  *  & $0.16\pm0.06$  \\ 
  Position Angle (rad)         & $4.23\pm11.14$   &  *  & $16.43\pm13.75$  \\
  Number of stars              &  $440$   &  $536$  &  $2550$      \\ \hline
\enddata
\tablecomments{~*: Measurements do not converge. $\sigma_{\mathrm{[Fe/H]}}$ and $\sigma_{\mathrm{Age}}$ correspond to the [Fe/H] and age intrinsic spreads, respectively. V${_r}$: Radial velocity. The correlation factor ($\rho$) indicates how age and metallicity are correlated and is estimated using Gaussian Mixture Models.}
\end{deluxetable*}
  
\subsection{subpopulations spatial distributions} \label{spatial_dist}

\begin{figure}
\centering
\includegraphics[width=240px]{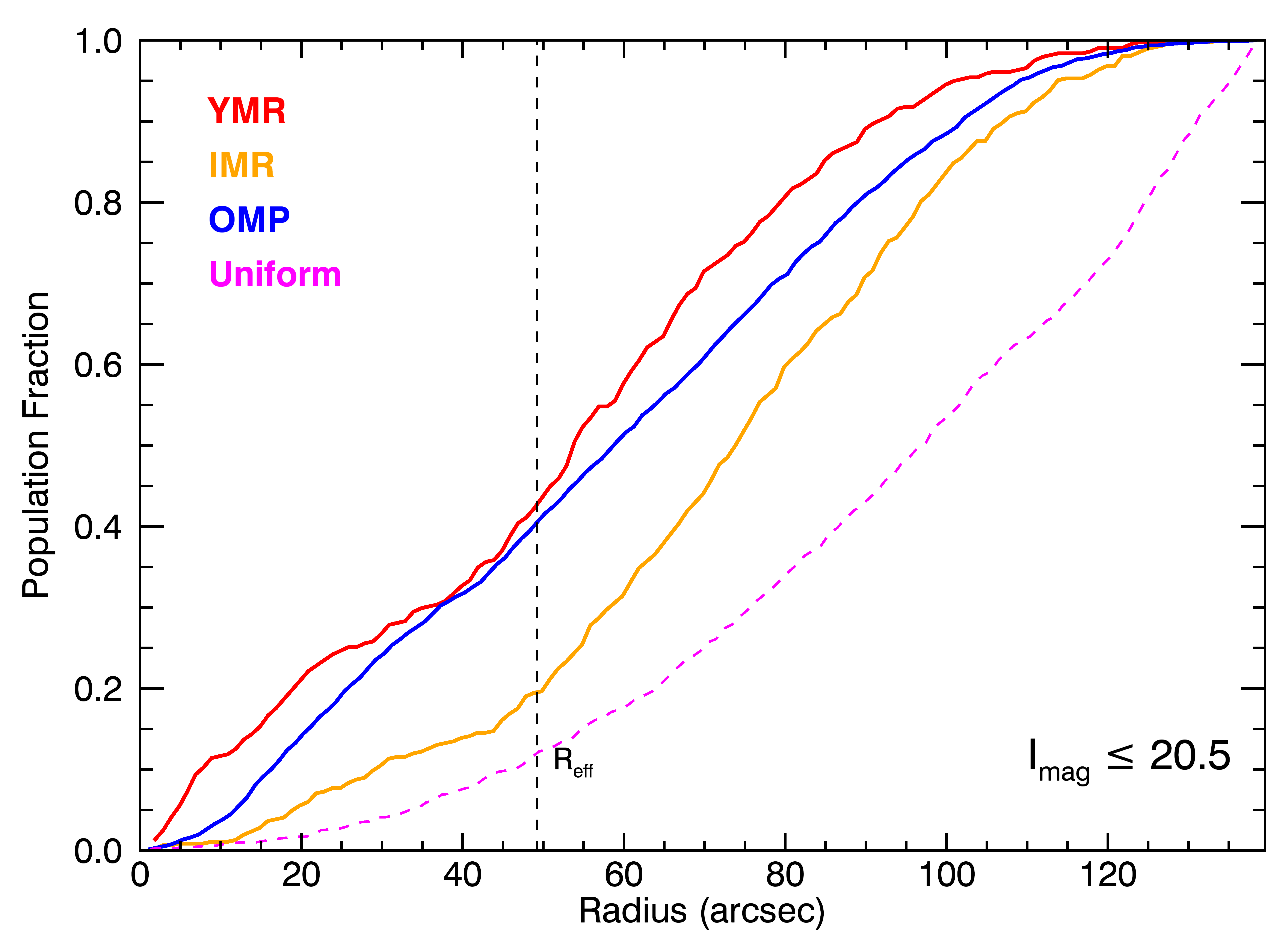}
\caption{Cumulative radial distribution for the three subpopulations for the member stars extracted with S/N$\geqslant10$ and magnitudes $\mathrm{I}\leqslant20.5$\,mag. YMR subpopulation in red ($2.2$\,Gyr,  \mbox{[Fe/H$=-0.04$}), IMR in orange (4.3\,Gyr, \mbox{[Fe/H]$=-0.29$}) and OMP ($12.2$\,Gyr, \mbox{[Fe/H]$=-1.41$}) in blue. The magenta dashed line describes a uniform stellar distribution. The vertical black dashed line shows the half-light radius of M54 \citep[R$_{\mathrm{HL}}=0\farcm82$,][2010 edition]{Harris1996}.}
\label{distr}
\end{figure}

We now analyze the spatial distribution of the three subpopulations in the Sgr dSph NSC defined in the previous section.
In Figure\,\ref{distr} we present the cumulative radial distribution for the three subpopulations: YMR (red), IMR (orange) and OMP (blue). We consider all the member stars extracted with signal to noise $\geqslant10$. In order to fairly compare the three subpopulations distribution in terms of completeness, we also constrain the sample to magnitudes $\mathrm{I}\leqslant20.5$\,mag.

In Figure\,\ref{distr} we observe that the YMR subpopulation is the most centrally concentrated of the three in this NSC.  The second highest central density is shown by the OMP subpopulation, which is the dominant in stellar number with over 2\,000 stars.  
Finally, the lowest central concentration is shown by the IMR subpopulation.  Comparing the distribution of these subpopulations with a uniform distribution (magenta dashed line), we observe that the stars in the IMR subpopulation are the least centrally concentrated in the most central 40". However, they are still not uniformly distributed, as we observed in Figure\,\ref{distr}, the distribution shows the stars in the IMR subpopulation are still significantly centrally peaked. This difference in the spatial distribution between the subpopulations suggests different origins. We discuss this in detail through Section\,\ref{discussion}.

It is important to consider that due to extreme central crowding and the limitations of our spatial resolution we are not able to extract all the stars that actually reside in this NSC. Improved spatial resolution is needed to more accurately derive the spatial distribution of the different stellar populations.

\subsection{2D morphology of subpopulations}

Using the coordinate information of the stars and the data FoV, we fit a 2-dimensional Plummer profile \citep{Plummer1911} to each of the subpopulations, optimizing the best profile using a MCMC algorithm \citep{Goodman2010, Foreman-Mackey2013}. We set the centroid, ellipticity, position angle (PA), and half-light radius as free parameters, thus obtaining the best estimates for each stellar subpopulation. 

The estimated parameters are summarized in Table\,\ref{table}. The likelihood for the YMR and OMP populations converge and give good estimates for the centroid, ellipticity and PA. However, for the IMR it does not converge due to the less centrally concentrated distribution that this subpopulation shows (see Section\,\ref{spatial_dist}).

\newpage

\section{Discussion} \label{discussion}

We show the systemic velocities of the three subpopulations in Table~\ref{table}.
The difference of $\sim1$ to $2\sigma$  between them could be a consequence of the number of stars in each sample, which is considerably higher for the OMP with $\sim$5 times more stars than the YMR and IMR subpopulations. This results in lower errors for the OMP. In spite of the slight difference the velocities are consistent.
The nearly identical systemic velocities derived from a high number of stars strongly supports the idea that all three subpopulations are spatially coincident, as also suggested in previous studies \citep[e.g.][]{Dacosta1995, Ibata1997, Monaco2005b, Bellazzini2008}. 
We also add another piece of evidence to this with their respective centroids (Table\,\ref{table}), showing that the subpopulations OMP and YMR are actually spatially coincident with the same centroid.
This strongly argues against a chance alignment in projection between the metal poor and metal rich populations \citep{Siegel2011}. 

\subsection{A possible merger remnant as the seed of the Sgr dSph NSC} \label{merger}

We find that the stars in the metal poor regime show a large spread in age, $\sigma_{\mathrm{age}}=0.9$\,Gyr, and iron content, $\sigma_{\mathrm{[Fe/H]}}=0.24$, a higher spread in comparison with the literature ($\sigma_{\mathrm{[Fe/H]}}=0.186$,  \citealt{Carretta2010b,Willman-Strader2012}). 
Large iron spreads in GCs can be explained by a scenario in which two GCs merge  \citep[][and references therein]{Gavagnin2016,Bekki2016,Khoperskov2018}.

\citet{Gavagnin2016} using N-body simulations studied the structural and kinematic signatures in the remnants of GC mergers with a sample of progenitors of different densities and masses. They pointed out that if two GCs were formed in the same dwarf galaxy (or molecular cloud) and display a low relative velocity they could merge, making the merging  scenario more likely in dwarf galaxies. They actually mention the Sgr dSph galaxy as a good candidate for this scenario to happen. \citet{Bellazzini2008} found the velocity dispersion profile is the close to flat with values of $\sim$10\,km~s$^{-1}$ in the innermost $80\farcm$.
\citet{Gavagnin2016} also include a kinematic analysis focused on the rotation of the remnants of merging GCs.  We will discuss this in a second paper focusing on the kinematic analysis for the different stellar subpopulations.

\citet{Bekki2016} used numerical simulations to suggest that GC mergers explain the existence of complexity in high-mass GCs, with multiple stellar populations and a large spread in metallicity, what the authors called "anomalous" GCs. 
Performing several tests, \citet{Bekki2016} found that the merger between GCs with masses greater than 3$\times$10$^5$\,M$_\odot$ will occur inevitably in the host dwarf galaxy with a mass in the range of M$_{dh}=3\times10^9$- $3\times10^{10}$\,M$_\odot$. 
This is due to a stronger dynamical friction effects on the GCs compared to the field stars in those shallow, low dispersion galaxies. The authors mention that if the clusters are massive enough, the timescale for the merger is just a few Gyr and it occurs before the total disruption of the dwarf galaxy due to the Galactic tidal field. They find that with a merger of at least two GCs it would be possible to observe GCs with an internal metallicity spread.
In the case of the Sgr dSph galaxy, the progenitor dark halo mass before infall to the Milky Way has been estimated with a lower limit of 6$\times10^{10}$\,M$_\odot$ by \citet{Gibbons2017}, in good agreement with \citet{Mucciarelli2017}. The mass estimated for M54 is \mbox{1.41$\pm0.02\times10^6$}\,M$_\odot$ \citep{Baumgardt2018}, consistent with mergers of GCs of $10^5$\,M$_\odot$. 

The merger scenario might explain the high mass, and the large iron and age spreads in the OMP subpopulation. The mass of the Sgr dSph progenitor would be conducive for the GCs to merge. From simulations, the infall of a massive GC to the center of Sgr dSph galaxy due to dynamical friction effects has been estimated to occur $\sim$5-9\,Gyr ago  \citep{Bellazzini2006a, Bellazzini2008}. The infall time is at most 3\,Gyr \citep{Bellazzini2008} for a cluster with M54's mass.
The spread in age and metallicity of the oldest, metal-poor stars is consistent with the scenario where multiple GCs were driven to the center of the Sgr dSph galaxy and merged, thus building up the metal-poor component of the NSC.

 \subsection{In situ formation in the Sgr dSph NSC: YMR subpopulation} \label{YMR}

From the metallicity and age characterization of our stellar sample we distinguish two subpopulations in the metal rich regime. 
We are able to separate them using the age-metallicity relation and CMD position of the stars. We observe that the youngest population - YMR - has the highest central density in the Sgr dSph NSC. The high metallicity, young age, and spatial distribution suggest that the formation of the YMR subpopulation formed {\em in situ} starting $\sim$3\,Gyr ago. 

Additional evidence for {\em in situ} formation of the YMR population is that it appears more flattened than the OMP population. We measure an ellipticity of 0.31$\pm$0.10 for the YMR component, while the OMP population has an ellipticity of 0.16$\pm$0.06 (Table\,\ref{table}).  Centrally concentrated and sometimes flattened young sub-components are commonly found in other NSCs in both early- and late-type galaxies \citep{Seth2006, Carson2015, Feldmeier-Krause2015, Nguyen2017}; given the young age of these populations, {\em in situ} formation is strongly favored, with the flattening suggestive of formation in a star forming gas disk.  
We note that the difference in ellipticity between the YMR and OMP subpopulations is small, given the measurement uncertainties. However, this difference becomes meaningful when we combine the ellipticity with the kinematics. In a second paper we will present the kinematic characterization of the subpopulations, adding a new diagnostic to address this issue. 

These young, flattened populations can survive despite being embedded in an older hotter population. \citet{mastrobuono-Battisti2013, Mastrobuono-Battisti2016} modelled the evolution of stellar disks in dense stellar clusters using $N$-body simulations, finding that even after several Gyr (around the Milky Way GCs age) the stars are still not fully mixed. Thus, both generations of stars show different distributions. They observed that the second population is concentrated at the center after 12\,Gyr with no relaxed spherical shape. 
The amount of rotation is also evidence for this model, where the second generation of stars is found to rotate faster than the first one. From our kinematic analysis, we found that the YMR subpopulation rotates faster than the OMP. We will present the evidence on this matter in our upcoming paper on this NSC kinematics.

To form the youngest and metal rich subpopulation enriched gas is needed, however no neutral gas has been detected in this galaxy. 
Using simulations, \citet{Tepper2018} found that the Sgr dSph galaxy might have lost all its gas after the second encounter with the Galactic disk. They found that two pericentric passages happened $\sim$2.8 and $\sim$1.3\,Gyr ago. The first produced a gas loss of 30 to 50\%, while the second would be responsible for the stripping of all the residual gas. The timescale of both encounters are consistent with the latest bursts of star formation seen in Figure~\ref{amr}, and were suggested based solely on the photometry by \citet{Siegel2007}.
This suggests that the YMR population may have formed by gas being driven into the nucleus during encounters with the Milky Way, supporting the {\em in situ} formation of this population.

The {\em in situ} formation of a new generation of stars could alternatively be explained for a massive NSC with a deep enough potential well to collect the enriched gas ejected from metal-rich and high-mass stars at larger radii that cools and sinks to the center \citep{Bailey1980}. This seems plausible given the characteristics of the stars in the IMR subpopulation and the relatively small fraction of YMR stars in comparison with the OMP.

 \subsection{Sgr dSph galaxy field stars in its NSC: IMR subpopulation} \label{IMR}
 
A different situation is observed for the IMR population in comparison to the YMR. These stars seem to be less centrally concentrated in the field with a wide spread in ages from 3 to 8\,Gyr. These findings suggest that the star formation in the Sgr dSph galaxy during this period was not particularly concentrated in the nucleus. The metallicity range of these stars is consistent with those found for stars close to the center of the galaxy \citep{Hasselquist2017}. 
In addition, comparing the IMR sample of our age-metallicity relation (Figure~\ref{amr}) with the one presented by \citet{Boer2015} for the Sagittarius bright stream we see a good agreement in both age (4-14\,Gyr) and metallicity ($-2.5<$[Fe/H]$<-0.1$). Thus, supporting our good recovery of the underlying chemical evolution of the host galaxy.

The spatial distribution of these two subpopulations - YMR and IMR - has also been reported by \citet{Mucciarelli2017}. Using a sample of 109 stars with [Fe/H]\,$\geqslant-1.0$ and a magnitude limit of I$_{\mathrm{mag}}=18$, the authors observed a shift in the metallicity  peak in the metal rich population at different projected distances from its center. At $0\farcm0<\mathrm{R}<2\farcm5$ the peak is at [Fe/H]$=-0.38$, changing to [Fe/H]$=-0.45$ at $2\farcm5<\mathrm{R}<5\farcm0$, noticing a metallicity gradient for this population.
In addition, the authors present the cumulative radial distribution of the two young populations where they found that the youngest subpopulation is more centrally concentrated than the intermediate age population. With this finding they suggest the youngest population is the dominant population in the Sgr dSph NSC, with the intermediate age one becoming more important at larger radii. Since our observations reach out to $\sim 2\farcm$ from the center of this NSC we are not able to see a change in the peak at larger radii. However, we also see this behavior between the two youngest metal rich subpopulations.

Our spectroscopic sample also includes stars in the blue plume (BP) region, which could be populated by young metal rich stars of [Fe/H]$=0.6$ and $0.1-0.8$\,Gyr \citep{Siegel2007} or blue straggler stars (BSS). However, we measure a wide range of metallicities for these stars, suggesting they are BSS instead of young stars, which would be expected to display a more homogeneous metallicity, e.g. as observed in the YMR subpopulation. \citet{Mucciarelli2017} found the BP and the intermediate age metal rich stars cumulative radial distributions were not distinguishable. Since this is a less dense environment in comparison to the center of the NSC, these BSS could be the product of mass transfer between binaries as has been found in other dSph  \citep[e.g.][]{Momany2007, Mapelli2007, Mapelli2009}.

As discussed in this section, the stars from the IMR subpopulation seem to have properties consistent with those at larger radii in the galaxy. This population shows a central concentration, shallower than the other two. With the current information we are not able to tell if the stars in the inner regions are actually dynamically bound to the NSC (with the YMR and OMP) or if they are part of the main body of the host galaxy.

\subsection{The Formation History of the Sgr dSph NSC}

From this work and previous studies of the populations in the Sgr dSph NSC, we can put together the story of this NSC. It starts with two or more massive GCs that eventually merge at the centre of the Sgr dSph galaxy, forming a massive nucleus consisting of old and metal poor stars with a large metallicity spread. 
The two encounters of the Sgr dSph galaxy with the Galactic disk occurred $\sim2.8$ and $\sim1.3$\,Gyr ago \citep{Tepper2018}, could have triggered two new episodes of star formation before the total stripping of the gas, creating in the first the YMR subpopulation. This results in a complex multi-population NSC.

This NSC is on its way to become a stripped nucleus considering the ongoing strong interaction between its host - the Sgr dSph galaxy - and the Milky Way. In fact, \citet{Bellazzini2008} suggested that this nucleus will probably end up as a compact remnant with two populations: old metal poor and young metal rich, with no signatures of the progenitor galaxy. This puts the Sgr dSph NSC in close company with the most massive cluster in the Milky Way, $\omega$Cen, which has long been considered a potential stripped nucleus \citep[e.g.][]{Lee1999,Carretta2010b}. 
$\omega$Cen presents a centrally concentrated disk-like component \citep[][see their Fig. 19, 20]{vandeVen2006}, very similar to the YMR subpopulation we detect in the Sgr dSph NSC. The age spread in $\omega$Cen \citep[at least 2\,Gyr,][]{Hilker2004, Villanova2014}, similar to the spread we see in the OMP suggests the merging of globular clusters early on in the nucleus of a progenitor galaxy \citep[e.g.][]{Bekki2016}.
Unlike $\omega$Cen, the stripping around the Sgr dSph NSC is ongoing, and thus we have the opportunity to understand the role mergers and stripping have played in creating the cluster we see today.

Given all the evidence we have presented in this paper, our suggestion to the community is to revert to the original naming and use "M54" in the same manner as it was given to the object upon its discovery by Charles Messier in 1778. The evidence consistently suggests M54 is not a normal metal-poor globular cluster but a complex NSC.

\section{Conclusion} \label{conclu}

In this work we present a rich sample of $\sim$6600 stellar spectra extracted from a mosaic of sixteen pointings of MUSE data on M54, the nuclear star cluster of the Sgr dSph galaxy, a dwarf galaxy currently being disrupted by the Milky Way.
Through radial velocity, metallicity ([Fe/H]) and age measurements we have characterized M54's stellar populations. We were able to detect at least three subpopulations with the same systemic velocity, differentiated by age and metallicity, where two of them have the same centroid.

The subpopulations we find are:  (i) YMR: young metal rich, with ages $2.2$\,Gyr and average metallicity \mbox{[Fe/H]$=-0.04$}, (ii) IMR: intermediate-age metal rich, with ages of $4.3$\,Gyr and metallicity \mbox{[Fe/H]$=-0.29$}, and (iii) OMP: old metal poor, with ages $12.2$\,Gyr and metallicity \mbox{[Fe/H]$=-1.41$}.

The existence of these three distinct subpopulations with the displayed differences in age and metallicity suggest the following conclusions:

\begin{itemize}

\item The stars in the OMP population have ages and metallicity consistent with it being assembled by two or more star clusters in-spiralling to the nucleus via dynamical friction. 

\item The YMR population is both more flattened and more centrally concentrated than the other two populations.  These features suggest {\em in situ} formation from enriched gas retained in the deep potential well of M54.  This young, centrally concentrated component is similar to features observed in other NSCs. We estimate the YMR subpopulation formation episode started around 3\,Gyr ago, consistent with the time of the first big encounter between Sgr dSph and the Milky Way, suggesting gas was channeled into the nucleus during this encounter. The youngest stars in the YMR population, $<$3\,Gyr, might be related to when Sgr dSph lost its gas during its ongoing interaction with the Milky Way. Alternatively, the YMR subpopulation could have formed from gas ejected from high-mass and metal-rich stars in the IMR subpopulation retained in the deep potential of the massive OMP subpopulation.

\item Our metallicity measurements for the IMR subpopulation are consistent with those for the field star population of the galaxy, including regions close to the center. This subpopulation shows the lowest central concentration in M54, but is still significantly centrally peaked. Additional information is needed to determine if these stars are actually dynamically bound to the NSC. \\

\end{itemize}

M54 is a unique test case. In this complex nucleus we find evidence for two processes that build up the NSC: (i) infall of two or more GCs, which merge to create a single high-mass cluster with a large metallicity spread, and (ii) in-situ star formation from enriched gas in the nucleus. In this case, the first scenario could be the key for the second to occur. This detailed formation history of the Sgr dSph NSC helps us understand the processes of NSC formation and the role of galaxy-galaxy interaction in this formation.

In a second paper we will present a kinematic analysis for the sample. We will include a kinematic analysis for the three age-metallicity subpopulations found in this work, giving additional insight on M54's formation. \\

\acknowledgments{
We thank the referee, Michele Bellazzini, for the constructive comments and suggestions that helped to improve this work.
SK gratefully acknowledges funding from a European Research Council consolidator grant (ERC-CoG-646928- Multi-Pop). 
RL acknowledges funding from a Natural Sciences and Engineering Research Council of Canada PDF award and RL and NN acknowledge support by Sonderforschungsbereich SFB 881 "The Milky Way System" (subprojects A7, A8 and B8) of the German Research Foundation (DFG).  MAC, RL and PB acknowledge support from DAAD PPP project number 57316058 "Finding and exploiting accreted star clusters in the Milky Way".
GvdV and LLW acknowledge funding from the European Research Council (ERC) under the European Union's Horizon 2020 research and innovation programme under grant agreement No 724857 (Consolidator Grant ArcheoDyn). This research has made use of NASA's Astrophysics Data System. We thank to Brian Kimmig for kindly provide his code "clumPy"\footnote{\url{https://github.com/bkimmig/clumpy}} for stellar cluster membership estimates.
We thank Alessandra Mastrobuono-Battisti and Iskren Georgiev for their helpful comments and discussions on this work.}

\bibliographystyle{yahapj}
\bibliography{mybib}{}

\appendix

\section{Comparing velocity and metallicity measurements for multiple stellar spectra} \label{repeated_stars}

\begin{figure*}
\centering
\includegraphics[width=\textwidth]{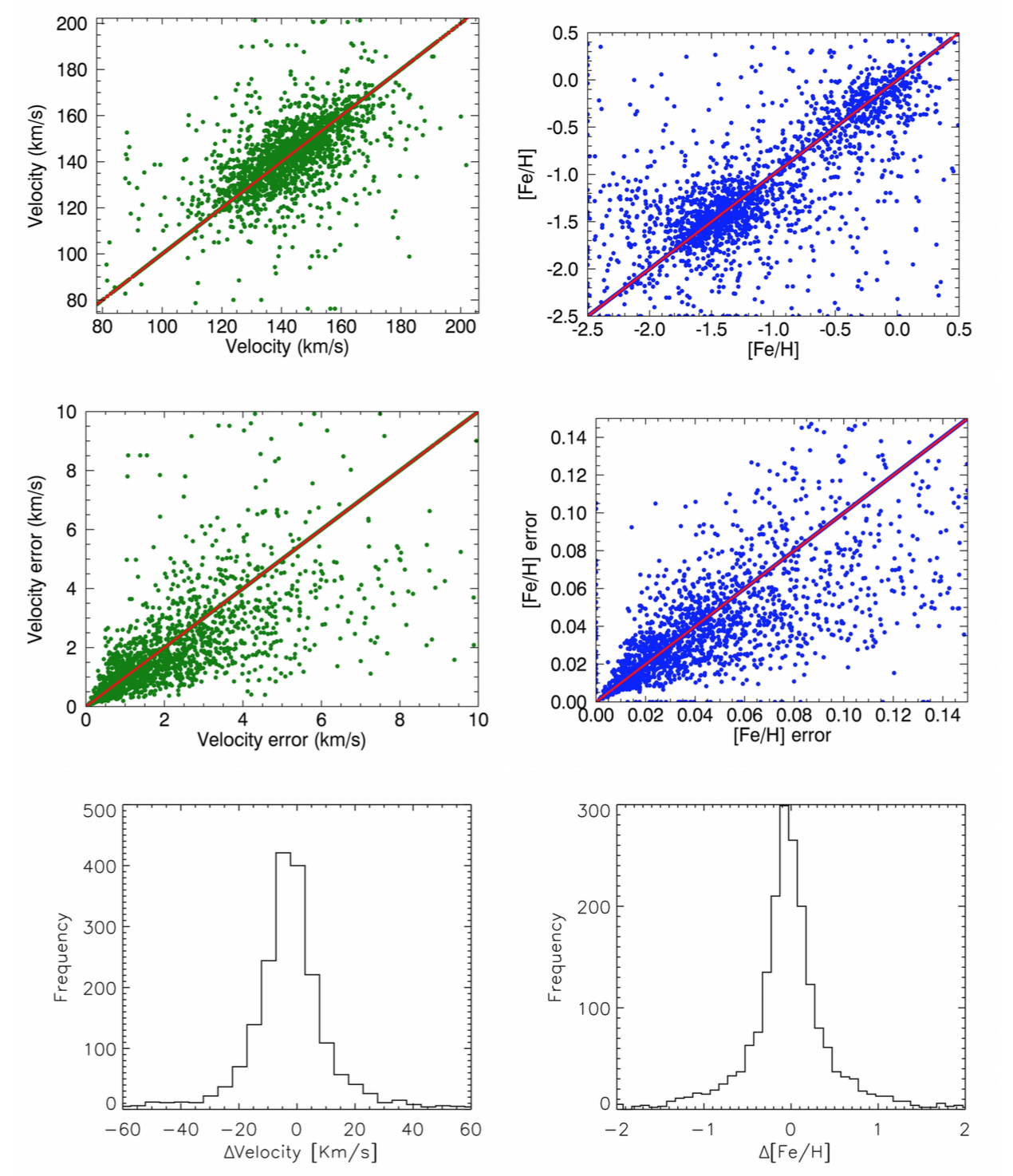}
\caption{The top left panel shows the velocity measurements correlation of the repeated stars. The middle left panel shows the correlation between the uncertainties. The bottom left panel shows the histogram of the difference in the measurements, with $\sigma=16.25$\,km~s$^{-1}$. The top right panel shows metallicity measurements correlation of the repeated stars. Their respective errors correlation is shown in the middle right panel. The bottom left panel shows the histogram of the difference in the measurements, with $\sigma=0.51$\,dex.
}
\label{star_comparison}
\end{figure*}

In Section\,\ref{stellar_extraction} we describe the stellar spectra extraction from the MUSE dataset. Due to overlaps between the neighboring pointings, we have from one up to four spectra for an individual star. To compare how the velocity, metallicity and their respective uncertainties are correlated in this appendix we include the comparison between the velocity estimates for the complete extracted stellar sample (see Section\,\ref{velocity_measurement}). The top left panel in Figure\,\ref{star_comparison} shows velocity versus velocity measurements of the repeated stars. We show the same for their corresponding errors in the middle left panel. The bottom left panel shows the histogram of the difference in the measurements. The estimated sigma is $\sigma=16.25$\,km~s$^{-1}$.

The right panels of Figure\,\ref{star_comparison} include the same comparison for the case of the metallicity measurements described in Section\,\ref{met_ulyss} for the same stellar sample. In the top right panel in Figure\,\ref{star_comparison} shows metallicity versus metallicity measurements of the repeated stars. The correlation between their respective errors is shown in the middle right panel. The bottom left panel shows the histogram of the difference in the measurements, with a sigma of $\sigma=0.51$\,dex.

Since brighter stars have higher signal-to-noise, to show the variation of the spread in dependency with magnitude, in Table\,\ref{table_spread} we include the standard deviation of the [Fe/H] and radial velocity measurements at different magnitude cuts.

The difference in the values for both parameters can be due to the different observing conditions during the acquisition of the 16 pointings (between June and July 2015). This produces variations in the quality of the spectra. In spite of this, we extracted good quality spectra with a reasonable signal-to-noise threshold of S/N=10. The dispersion we observe between repeated stars is acceptable and does not affect our main results.  

\begin{deluxetable}{c|ccc}[ht!]
\tablecaption{Radial velocity and [Fe/H] spread for repeated stars. \label{table_spread}}
\tablehead{
\colhead{Imag range}  &  \colhead{$\sigma_{\mathrm{V_r}}$ (km~s$^{-1}$)} & \colhead{$\sigma_{\mathrm{[Fe/H]}}$} & \colhead{Number of stars} }
\startdata
  $[13 - 14)$ &   $5.84$  & $0.19$  &  $55$      \\
  $[14 - 15)$ &  $12.75$  & $0.40$  &  $135$     \\
  $[15 - 16)$ &  $13.12$  & $0.51$  &  $235$     \\
  $[16 - 17)$ &  $16.83$  & $0.54$  &  $425$     \\
  $[17 - 18)$ &  $17.78$  & $0.54$  &  $430$     \\
  $[18 - 19)$ &  $12.30$  & $0.47$  &  $324$     \\
  $[19 - 20)$ &  $16.87$  & $0.50$  &  $236$     \\
  $[20 - 21)$ &  $29.51$  & $0.58$  &  $68$      \\
  $[21 - 22)$ &  $30.75$  & $0.42$  &  $6$       \\
  $[13 - 22]$ &  $16.25$  & $0.51$  &  $1\,920$   \\ \hline
\enddata
\tablecomments{~$\sigma$ correspond to the spread given by the standard deviation of the measurements from the repeated stars at different magnitude cuts.}
\end{deluxetable}
  
\vspace{0.5cm}

\section{Metallicity comparison with previous works} \label{feh_comparison}

\begin{figure*}
\centering
\includegraphics[width=500px]{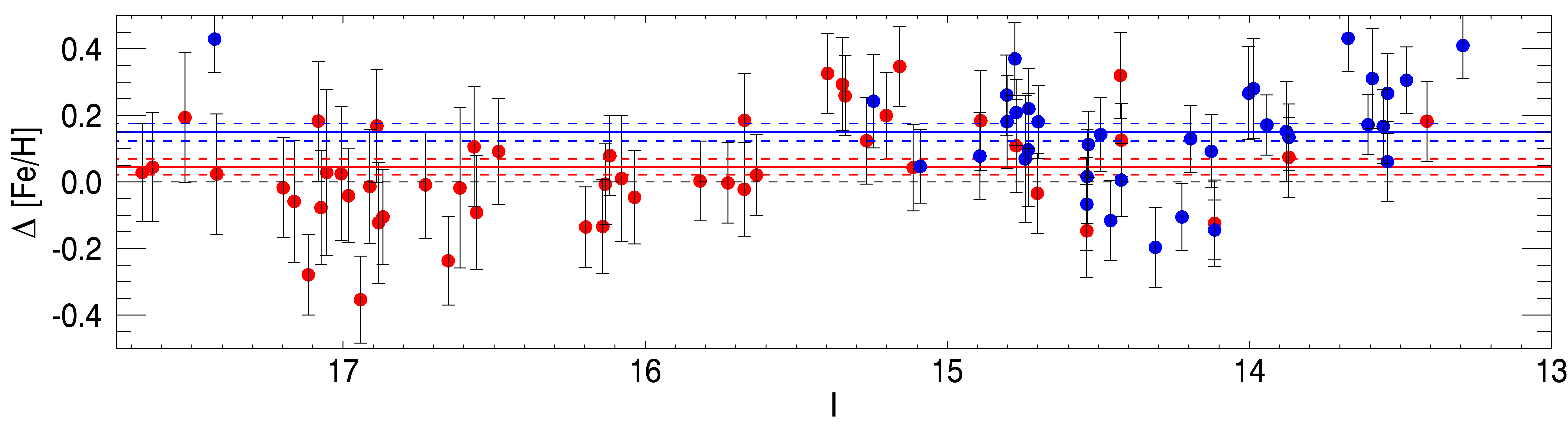}\\
\includegraphics[width=350px]{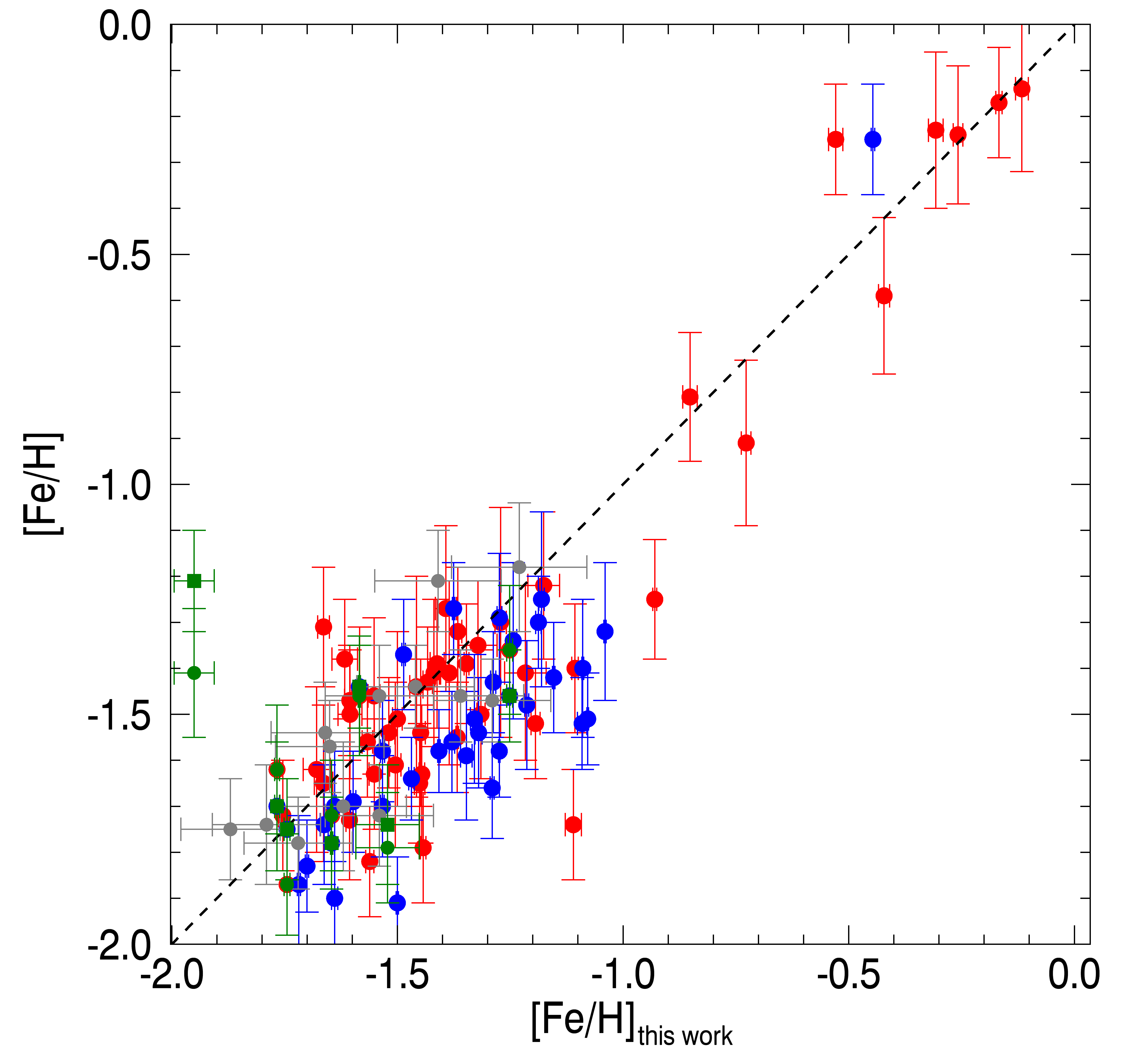}
\caption{\textbf{Top}: Difference in metallicity [Fe/H] between this and other works \textbf{for stars  with $13\leqslant\mathrm{I}\leqslant17.75$\,mag}. The 51 red circles show the $\Delta$[Fe/H] between this work and \citet{Mucciarelli2017}, and the 36 blue circles for \citet{Carretta2010a}.
The mean $\Delta$[Fe/H] between this work and \citet{Mucciarelli2017} is 0.05$\pm$0.02\,dex, with $\sigma=0.17$. For $\Delta$[Fe/H] between this work and \citet{Carretta2010a}, the mean is 0.15$\pm$0.03\,dex, with $\sigma=0.16$.
The continuous red and blue lines show the mean value of the respective $\Delta$[Fe/H] estimates, while the same color dashed lines show the area confined by the mean error. The dashed black line shows a $\Delta$[Fe/H]=0. 
\textbf{Bottom}: Relation between [Fe/H] estimates from this work versus other works: red circles for \citet{Mucciarelli2017} (51 stars), and blue circles  for \citet{Carretta2010a} (36 stars).  The gray circles represent the relation of 13 common stars between \citet{Carretta2010a} and \citet{Mucciarelli2017}. The 8 common stars between the three samples are represented as green circles when comparing this work with \citet{Mucciarelli2017} and green squares with \citet{Carretta2010a}.  The dashed black line shows a one to one relation. }
\label{feh_comparison_figure}
\end{figure*}

The FoV of our MUSE data partially overlaps with those in \citet{Carretta2010a} and \citet{Mucciarelli2017}. We cross matched the samples using coordinate information for the stars with $\mathrm{I}\leqslant17.75$\,mag, since the S/N of the spectra is larger and warrants reliable metallicity measurements. In the top panel of Figure \ref{feh_comparison_figure} we present the difference in [Fe/H] ($\Delta$[Fe/H]) of the 51 stars that are in common between this work and \citet{Mucciarelli2017} (red circles), and the 36 with \citet{Carretta2010a}  (blue circles). The dashed black line shows a $\Delta$[Fe/H]=0.
For $\Delta$[Fe/H] between this work and \citet{Mucciarelli2017}, we obtain a mean value of 0.05$\pm$0.02\,dex, with a standard deviation of $\sigma=0.17$. For $\Delta$[Fe/H] between this work and \citet{Carretta2010a}, a mean value of 0.15$\pm$0.03\,dex, with a standard deviation of $\sigma=0.16$.
With the respective colors, we include in the figure the $\Delta$[Fe/H] mean values as continuous lines, the dashed lines show the area confined by the error of the mean. In this analysis we observe a good agreement between our metallicity measurements with previous studies, which use different methods and in the case of \citet{Carretta2010a}, high-resolution spectra.
 
In the bottom panel of Figure \ref{feh_comparison_figure} we show the relation between the metallicity values from this work versus \citet{Mucciarelli2017} (51 stars, red circles) and \citet{Carretta2010a} (36 stars, blue  circles). In addition, we include the 13 stars in common between \citet{Carretta2010a} and  \citet{Mucciarelli2017} as gray circles. The cross match between these 13 stars and our sample gives 8 stars in common. These stars are included in green circles when comparing this work  with \citet{Mucciarelli2017} and green squares when comparing with \citet{Carretta2010a}. 

\vspace{0.5cm}

\section{CaII triplet metallicity} \label{caii}

For the metallicity estimates, we use two independent methods to test the robustness of our results. 
The second method we use to estimate the metallicity is based on the calcium triplet lines (CaII triplet) at 8498,8542,8662\,\AA\, (see, e.g.  \citealt{Starkenburg2010, Carrera2013}). We use the calibration presented by \citet{Carrera2013}, obtained using the CaII triplet and calibration measurements obtained from hundreds of red giant branch star observations.

These non-linear calibrations are specifically tailored towards the metal poor regime, reaching reliable estimates of metallicity for stars between $-4.0<$[Fe/H]$<0.5$.
CaII triplet lines are strong on the red giant branch. But their strength depends on the joint effects from metallicity, temperature and surface gravity. Therefore, to obtain just the effect due to the metallicity, the authors removed these last two contributions using luminosity indicators in order to obtain the following relation (equation (2) in \citealt{Carrera2013}):

\begin{equation}
\begin{split}
\mathrm{[Fe/H]}\,&=\,a\,+\,b\,\times\,\mathrm{Mag}\,+\,c\,\times\Sigma \mathrm{Ca}\,+\,d\,\times \Sigma \mathrm{Ca}^{−1.5}\,\\ 
& +\,e\times\,\Sigma\,\mathrm{Ca}\,\times\,\mathrm{Mag}
\label{carrera_calibration}
\end{split}
\end{equation}

We use Mag - the luminosity indicator - as $V\,-\,V_{\mathrm{HB}}$, where $V_{\mathrm{HB}}\,=\,18.16$ \citep{Harris1996} is the average magnitude of the horizontal branch. The coefficient values according to \citet{Carrera2013} are as follows: $a\,=\,-3.45$, $b\,=\,0.11$, $c\,=\,0.44$, $d\,=\,-0.65$, $e\,=\,0.03$. $\Sigma \mathrm{Ca}$ corresponds to the sum of the equivalent width  ($\Sigma$\,EW) of the three CaII lines. For additional details see \citet{Carrera2013}.

\begin{figure*}
\centering
\includegraphics[width=300px]{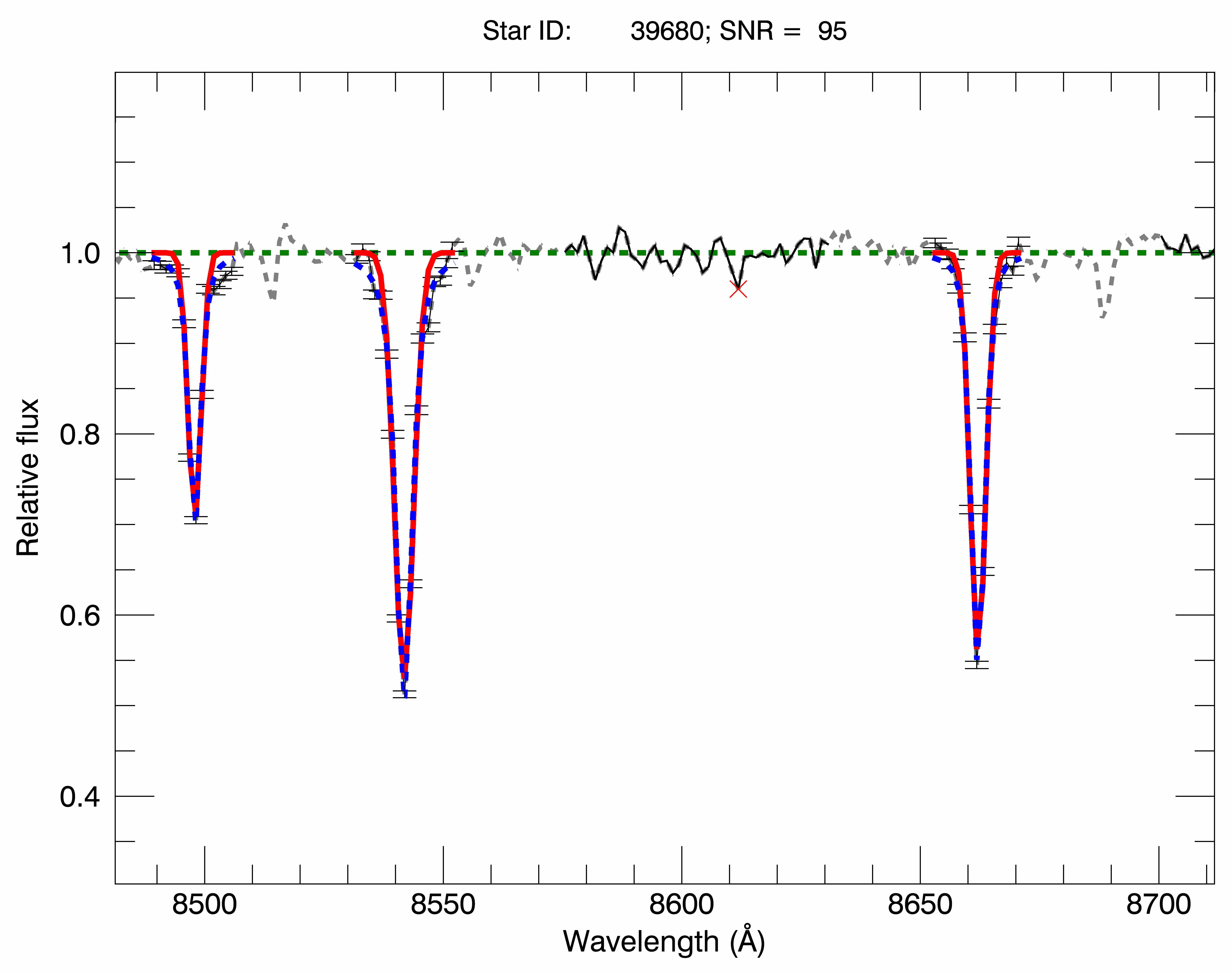}\\
\includegraphics[width=300px]{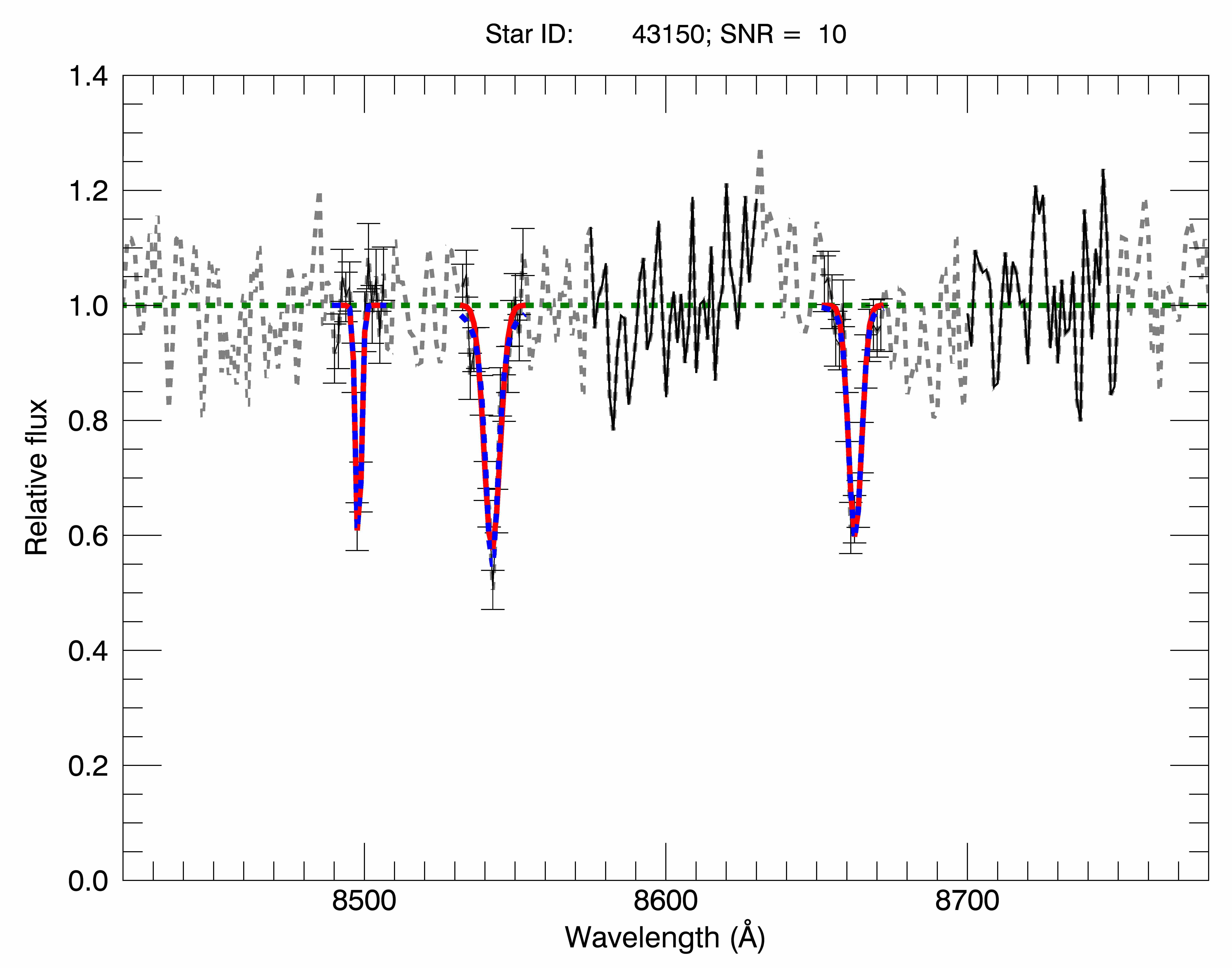}
\caption{Fitting examples of Voigt (dashed red line) profiles to the CaII triplet lines for two spectra. \textbf{Top}: Example of high signal to noise  spectrum (S/N$=95$). \textbf{Bottom}: Example of low signal to noise  spectrum (S/N$=10$). }
\label{fitting_caii}
\end{figure*}

We applied the Doppler correction to the spectra using the radial velocity values of each star and performed a continuum normalization. We fit a Gauss and Voigt profile to estimate the EW of the CaII triplet lines. We use the spectral ranges for these lines defined by \citet{Armandroff1988}. We find that Voigt profiles provided better fits than Gaussian profiles for high S/N, therefore we have used Voigt profiles fits for estimating the metallicity. In Figure\,\ref{fitting_caii} are shown fitting examples for two stars with different signal to noise values (S/N$=10$ and S/N$=95$). The best-fit Voigt profiles is overplotted as a dashed red line.

\begin{figure}
\centering
\includegraphics[width=250px]{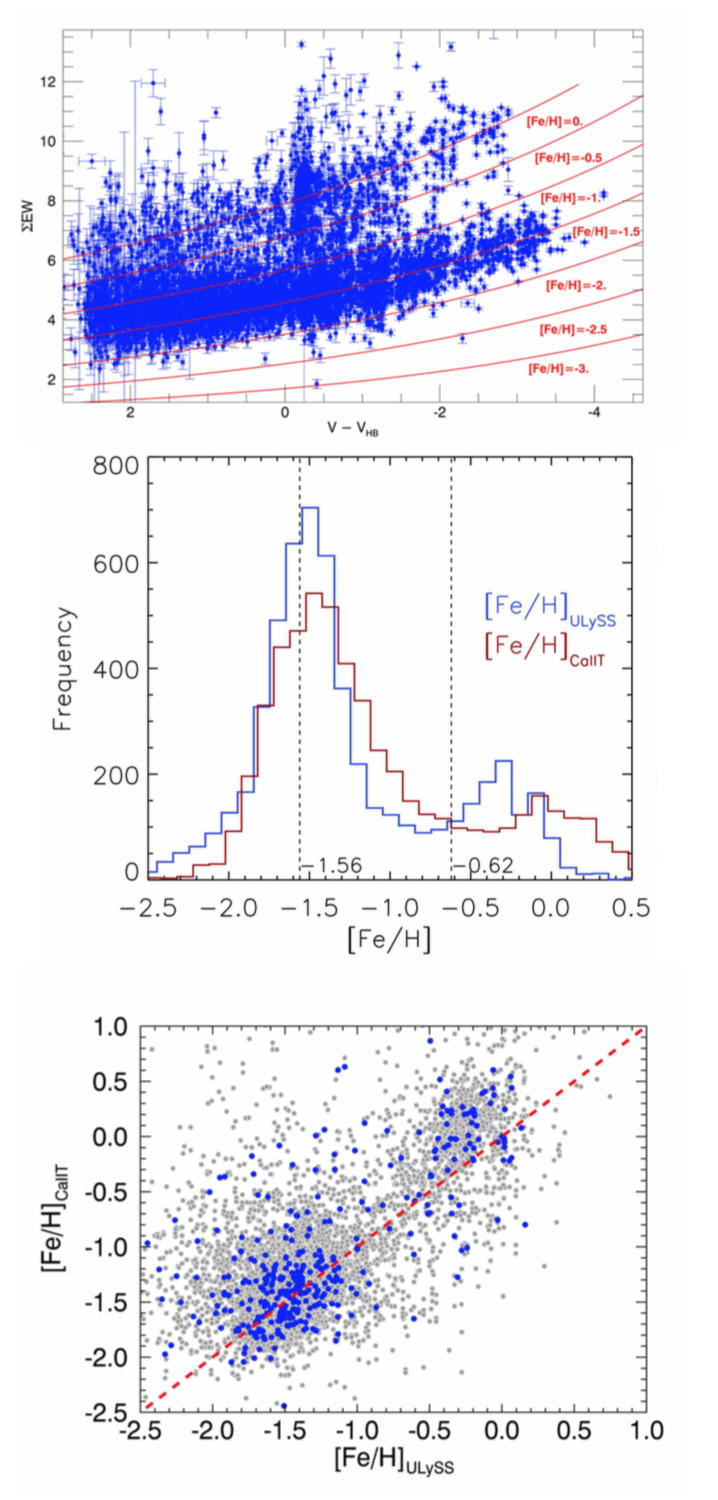}\\
\caption{Top panel: Equivalent width sum of the three CaII triplet lines versus V$-$V$_{\mathrm{HB}}$. Overplotted in red lines are the metallicity values for the Carrera calibration using the spectral ranges of the CaII triplet lines defined by \citet{Armandroff1988}.
Middle panel: Metallicity histogram of the values  obtained  with the CaII triplet lines method using the calibration published by \citet{Carrera2013} (red) and those obtained performing full spectral fitting using ULySS \citep{Koleva2009} for the same stellar sample (blue). There are fewer stars in the first method since it is restricted to RGB stars ($\sim$4\,000). From left to right the vertical black lines show the average metallicity for the metal poor and metal rich stars estimated by \citet{Carretta2010a}.
Bottom panel: Correlation of the metallicity values obtained with the two methods, for spectra with signal-to-noise $>60$ in blue and $<60$ in gray.}
\label{caii_met}
\end{figure}

Using the $\Sigma$\,EW values from the Voigt fitting in the calibration in equation\,\ref{carrera_calibration} we obtain metallicity values from \mbox{$-2.9<$[Fe/H]$<0.5$}.
Since this calibration is limited to RGB stars, for this procedure we consider a cut in \mbox{V$-$I$=0.7$}, selecting 4\,051 stars for the metallicity estimates in the sample of members of the Sgr dSph NSC. In the top panel of Figure\,\ref{caii_met} is shown the $\Sigma$\,EW versus V$-$V$_{\mathrm{HB}}$, overplotted in red lines are the metallicity values for this calibration.
The middle panel shows the metallicity histogram of the metallicity values obtained with this method (red). For comparison, we have overplotted the metallicity histogram of the estimates obtained performing full spectral fitting using ULySS \citep{Koleva2009} (blue) for the same sample of RGB stars (see Section\,\ref{met_ulyss}). The vertical black line on the left shows the average metallicity for the metal poor population \mbox{([Fe/H]$=-1.559\pm0.021$\,dex)}, estimated by \citet{Carretta2010a} using neutral Fe lines of 76 stars. The vertical black line to the right shows the average metallicity for 25 metal rich stars  \mbox{([Fe/H]$=-0.622\pm0.068$\,dex)} estimated by the same authors.
We see consistent results between our metallicity measurements using full spectral fitting with ULySS \citep{Koleva2009} and the measurement by \citet{Carretta2010a} for the metal poor stars. For the metal rich population we see a difference in the average metallicity, our peak a more metal rich ([Fe/H]$\sim-0.3$\,dex). We give a possible explanation for this difference in Section\,\ref{3pop}.
The bottom panel shows the correlation of the metallicity values between the two methods where we see a bigger difference in the high metallicity regime. The estimates are plotted in blue for the spectra with S/N$>60$ and in gray for S/N$<60$.

\end{document}